\documentclass[aps,prd,onecolumn,notitlepage,nofootinbib,eqsecnum,superscriptaddress]{revtex4-2}
\usepackage[T1]{fontenc}
\usepackage{amsmath,amssymb,bm,mathrsfs,booktabs,array,longtable,graphicx,mathtools,xcolor}
\usepackage[colorlinks=true,linkcolor=blue,citecolor=blue,urlcolor=blue]{hyperref}
\allowdisplaybreaks[2]
\begin{document}
\title{Parity-violating scalar-tensor theory revisited:\\ complete basis up to $d=4$ and Hamiltonian analysis}

\author{Yun-Chuan Zhang}
\affiliation{School of Physics, Sun Yat-sen University, Guangzhou 510275, China}

\author{Yu-Min Hu}
\affiliation{School of Physics and Electronic Information, Gannan Normal University, Ganzhou 341000, China}

\author{Xian Gao}
\email[Corresponding author: ]{gaoxian@mail.sysu.edu.cn}
\affiliation{School of Physics, Sun Yat-sen University, Guangzhou 510275, China}
\affiliation{Guangdong Provincial Key Laboratory of Quantum Metrology and Sensing,\\ Sun Yat-sen University, Zhuhai 519082, China}

\date{November 5, 2026}
\begin{abstract}
We construct complete bases of parity-violating generally covariant scalar-tensor (GST) and spatially covariant gravity (SCG) monomials up to derivative grade $d=4$. Including all derivative partitions, each description contains one $d=3$ and thirteen $d=4$ linearly independent monomials. At the level of Lagrangians, we find six integration-by-parts relations that reduce polynomial-type Lagrangians to combinations of one $d=3$ and seven $d=4$ monomials with general coefficient functions. We derive the correspondence between the two bases, retaining general coefficients and total derivatives, and establish their independence under integrations by parts. Among the seven $d=4$ SCG monomials, five contain neither the lapse velocity nor the time derivative of the extrinsic curvature. For the two remaining monomials, we perform a Hamiltonian constraint analysis with a standard Einstein-Hilbert kinetic term and a lapse-dependent potential. On the regular branches considered, the lapse-velocity term propagates four degrees of freedom (DOFs), while the extrinsic-curvature-velocity term propagates six DOFs, whether present alone or together with the lapse-velocity term. These results provide a complete basis for constructing parity-violating scalar-tensor theories with three DOFs up to $d=4$ within this setting, with the parity-violating Lagrangian spanned by six GST polynomials: one $d=3$ and five $d=4$ polynomials.
\end{abstract}

\maketitle
%\tableofcontents

%%
\section{Introduction}

A key problem in constructing scalar-tensor theories is to control the number of propagating degrees of freedom in the presence of higher order derivatives. Galileon interactions and their covariant extension illustrate this problem \cite{Nicolis:2008in,Deffayet:2009wt}. Second-order field equations lead to Horndeski theory and its generalized-Galileon formulation \cite{Horndeski:1974wa,Deffayet:2011gz,Kobayashi:2011nu}, while beyond-Horndeski and degenerate higher-order scalar-tensor (DHOST) theories remove unwanted canonical pairs through constraints \cite{Gleyzes:2014dya,Langlois:2015cwa,BenAchour:2016fzp,Langlois:2015skt}. Complementary constructions use probe-brane geometry \cite{deRham:2010eu,Goon:2011qf} and disformal transformations \cite{Bekenstein:1992pj,Bettoni:2013diz,Zumalacarregui:2013pma,BenAchour:2016cay}. Invertibility of the latter is essential when identifying solution spaces and degrees of freedom \cite{Takahashi:2017zgr}.

Spatially covariant gravity (SCG) separates temporal and spatial derivatives at the action level \cite{Gao:2014soa,Gao:2014fra}. Its building blocks include the lapse, spatial metric, extrinsic and intrinsic curvatures, and their spatial derivatives. Without time-reparametrization invariance, a regular theory generically contains one scalar mode in addition to two tensor modes. A lapse velocity requires further degeneracy and consistency conditions to avoid an additional mode \cite{Gao:2018znj,Gao:2019lpz}. The SCG framework also underlies the effective theories of inflation and dark energy \cite{Cheung:2007st,Gubitosi:2012hu,Gleyzes:2014rba}, while Ho\v{r}ava gravity uses temporal--spatial separation for anisotropic ultraviolet scaling \cite{Horava:2009uw,Blas:2009qj}. Massive gravity and solid inflation provide other symmetry-breaking constructions with different field content \cite{deRham:2010kj,Endlich:2012pz}.

For a timelike scalar gradient, unitary gauge $t=\phi$ relates generally covariant scalar-tensor (GST) and SCG actions. Systematic GST monomial classifications \cite{Gao:2020juc}, gauge-fixing and gauge-recovery relations \cite{Gao:2020yzr}, linear monomial mappings \cite{Gao:2020qxy}, and covariant $3+1$ methods \cite{Hu:2021bbo} make this correspondence explicit. Unitary-gauge degeneracy is weaker than covariant DHOST degeneracy. In the unitary-degenerate higher-order scalar-tensor (U-DHOST) theories studied in Refs.~\cite{DeFelice:2018ewo,DeFelice:2021hps}, an apparent extra mode obeys an elliptic spatial equation fixed by boundary conditions. Mode counts therefore require a specified timelike-gradient branch and boundary problem. SCG also admits scalar elimination through Hamiltonian conditions \cite{Gao:2019twq}, perturbative conditions \cite{Hu:2021yaq,Yu:2026fgn}, auxiliary constraints \cite{Yao:2020tur,Yao:2023qjd}, or nondynamical scalars \cite{Gao:2018izs,Wang:2024hfd,Zhu:2025bcu}. Extensions include nonmetricity \cite{Yu:2024drx}, multiple scalar-defined foliations \cite{Yu:2024sed,Chen:2026hhl}, and spatially covariant vector theories \cite{Li:2025qbv}.

SCG theories have been widely applied in the study of cosmology and gravitational waves, including cosmological perturbations and disformal transformations \cite{Fujita:2015ymn}, dynamical-lapse perturbations \cite{Zhu:2022caa}, and gravitational-wave propagation and polarization \cite{Gao:2019liu,Zhu:2022dfq,Zhu:2022uoq}. Related effective theories describe black-hole perturbations and tidal responses \cite{Mukohyama:2022skk,Barura:2024uog}. 

In particular, parity violation can distinguish the circular gravitational-wave polarizations. Chern--Simons gravity \cite{Jackiw:2003pm}, its inflationary realizations and scalar-tensor extensions \cite{Satoh:2007gn,Satoh:2010ep,Qiao:2019hkz}, and fluctuation effective theories \cite{Creminelli:2014wna,Cannone:2015rra} provide representative mechanisms. Their CMB signatures \cite{Lue:1998mq,Saito:2007kt,Contaldi:2008yz} must be distinguished from polarization rotation \cite{Gluscevic:2010vv}. Compact-binary detections \cite{LIGOScientific:2016aoc,LIGOScientific:2017vwq} enable tests through emission and propagation \cite{Alexander:2017jmt,Qiao:2019wsh,Zhao:2019xmm,Jenks:2023pmk}, including speed and birefringence constraints \cite{Nishizawa:2018srh,Wang:2020cub,Wang:2021gqm,Okounkova:2021xjv,Zhu:2023rrx}, stochastic-background searches \cite{Callister:2023tws}, and detector forecasts \cite{Zhang:2025kcw}.

Further applications concern scalar-induced waves \cite{Zhang:2022xmm,Feng:2023veu,Zhang:2023scq}, primordial spectra, trispectra, and circular polarization \cite{Li:2024fxy,Moretti:2024fzb,Feng:2026pcs}, and rotating black holes \cite{Yu:2025fcv}. Parity-odd interactions also arise from torsion \cite{Li:2020xjt,Li:2021wij}, nonmetricity \cite{Conroy:2019ibo}, and independent Palatini connections \cite{Sulantay:2022sag}. These applications motivate an operator classification with explicit dynamical assumptions. In metric scalar-tensor theory, Ref.~\cite{Crisostomi:2017ugk} constructs chiral extensions of Chern--Simons gravity and derives coefficient conditions removing higher temporal derivatives in unitary gauge. Within that ansatz these conditions select viable combinations, but do not establish a complete independent basis including higher scalar and curvature derivatives or coefficient-dependent integrations by parts.

Reference~\cite{Hu:2024hzo} addressed this classification through the SCG--GST correspondence, extending the parity-odd SCG monomials of Ref.~\cite{Gao:2019liu} to include the lapse velocity and the Lie derivative of the extrinsic curvature through derivative grade $d=4$. Here the integer $d$ is the total number of derivatives in the corresponding SCG expressions. The resulting list contained one $d=3$ and eight $d=4$ monomials. The subset without higher order time derivatives was mapped to generally covariant combinations in order to derive the ghost-free GST theories. This construction was recently extended to $d=5$ in Ref.~\cite{Song:2026ajt}, which also identified an algebraic relation among the three $KK\mathrm{D}K$ contractions in Ref.~\cite{Hu:2024hzo}. This relation implies that only two contractions are independent, reducing the number of independent $d=4$ SCG monomials to seven and the subset without higher order time derivatives to five. 
Nevertheless, a thorough investigation of independent monomials and thus the complete basis of polynomials in both SCG and GST frameworks is still lacking. In particular, this requires a careful distinction between algebraic independence and integration-by-parts relations. 
Moreover, whether the two SCG monomials with higher order time derivatives propagate extra degrees of freedom (DOFs) has not been examined.

This work addresses these questions. First, we complete the construction and classification of parity-violating (PV) GST and SCG monomials through $d=4$, including higher-derivative building blocks and products of monomials. 
We determine the complete basis of monomials in the sense of algebraic independence.
We then examine the integration-by-parts relations among these monomials, with arbitrary coefficient functions.
Second, we perform a Hamiltonian constraint analysis to determine the number of DOFs of the two SCG monomials involving higher order time derivatives. We use the standard Einstein-Hilbert kinetic completion with a lapse-dependent potential. 
This analysis determines whether these two monomials propagate extra DOFs beyond the two tensor and one scalar DOFs in an ordinary SCG theory without higher time derivatives. Through the SCG-GST correspondence, we then identify, up to $d=4$, the most general parity-violating scalar-tensor theory without extra DOFs within this setting.

This paper is organized as follows. Sec.~\ref{sec:basis} constructs the complete GST and SCG bases, establishes their correspondence and independence under integrations by parts, and identifies the subsets without higher time derivatives. Sec.~\ref{sec:F} presents the Hamiltonian constraint analysis of the lapse-velocity interaction. Sec.~\ref{sec:L} analyzes the extrinsic-curvature-velocity interaction and determines its degrees of freedom. Sec.~\ref{sec:both} examines both interactions together and whether their combination removes the additional degrees of freedom. Sec.~\ref{sec:conclusion} summarizes our results and discusses their implications.

Notations: We use spacetime indices $a,b,c,\ldots$ and spatial indices $i,j,k,\ldots$. We adopt the signature $(-+++)$ and the curvature convention $[\nabla_a,\nabla_b]w_c=R_{cdab}w^d$ and $R_{ab}=g^{cd}R_{cadb}$.
The four-dimensional volume form obeys $\varepsilon_{0123}=1$ in an oriented orthonormal frame.

\section{Complete basis of parity-violating polynomials}
\label{sec:basis}

In this section, we construct complete bases of PV GST and SCG monomials through $d=4$, distinguishing pointwise linear independence from independence under integrations by parts. Their explicit correspondence identifies the five independent $d=4$ combinations without higher time derivatives.

\subsection{Complete basis of parity-violating GST polynomials}
\label{subsec:GSTbasis}

\subsubsection{Linear space of GST polynomials}

The GST monomials up to $d=4$ have been investigated and classified in Ref.~\cite{Gao:2020juc}. Here we revisit the analysis by focusing on the independence of parity-violating monomials.
By definition,  a GST monomial is a scalar built of the scalar field, the curvature tensor and their covariant derivatives.
Following Ref.~\cite{Gao:2020juc}, an operator containing $c_n$ factors of $\nabla^nR_{abcd}$ and $d_m$ factors of $\nabla^m\phi$ has
\begin{equation}
d \coloneqq \sum_{n = 0}(n+2)c_n+\sum_{m= 1}(m-1)d_m.
\label{grading}
\end{equation}
We identify $\phi_a \equiv \nabla_a\phi$ as of $d=0$, which is also consistent with the above definition of $d$.
As a result, the integer $d$ corresponds to the total number of derivatives in a monomial modulo the first derivative $\phi_{a}$.

We follow the classification and notation of Ref.~\cite{Gao:2020juc}. The superscript $(c_0;d_2,d_3)$ specifies the numbers of $R_{abcd}$, $\phi_{ab}\equiv \nabla_a\nabla_b\phi$, and $\nabla_a\phi_{bc}$, respectively. Thus $d=2c_0+d_2+2d_3$. The symbols $E_n^{(c_0;d_2,d_3)}$ and $F_n^{(c_0;d_2,d_3)}$ denote parity-even and parity-odd monomials, with $n$ distinguishing monomials in the same class. 

Ref.~\cite{Gao:2020juc} exhaustively constructs the monomials built from these three building blocks. At $d=3$, the retained classes are $(0;3,0)$, $(0;1,1)$, and $(1;1,0)$. At $d=4$, they are $(0;4,0)$, $(0;2,1)$, $(0;0,2)$, $(1;2,0)$, $(2;0,0)$, and $(1;0,1)$. Our first task is to examine the linear independence of this generating set, without discarding divergences. The classes involving derivatives of curvature, which are excluded at this stage of the construction in Ref.~\cite{Gao:2020juc}, will be included separately below.

There are no PV scalars at $d\leq2$. At $d=3$, the two nonzero classes obey $F_1^{(0;1,1)}=-\frac12F_1^{(1;1,0)}$. Thus the unique PV monomial at $d=3$ is
\begin{equation}
\mathcal G_0=F_1^{(1;1,0)}=\sigma^{-3}\varepsilon_{abcd}R_{ef}{}^{cd}\phi^a\phi^e\phi^{bf}.
\label{Gzero}
\end{equation}
Here and throughout this paper, we denote $X=-\frac12\phi_a\phi^a>0$ and $\sigma=\sqrt{2X}$. We consider finite local tensor polynomials in higher derivatives, with the displayed powers of $\sigma$ and arbitrary smooth coefficients $f(\phi,X)$.

At $d=4$, for our purpose, we select the following seven monomials:
\begin{align}
\mathcal G_1&=F_2^{(1;2,0)}=\sigma^{-4}\varepsilon_{abcd}R_{ef}{}^{cd}\phi^a\phi^e\phi_m{}^b\phi^{fm},\label{Gbasis-1}
\\
\mathcal G_2&=F_5^{(1;2,0)}=\sigma^{-4}\varepsilon_{abcd}R_{ef}{}^{cm}\phi^a\phi^e\phi^{bf}\phi_m{}^d,\label{Gbasis-2}
\\
\mathcal G_3&=F_6^{(1;2,0)}=\sigma^{-4}\varepsilon_{abcd}R^{ae}\phi^b\phi^f\phi_e{}^c\phi_f{}^d,\label{Gbasis-3}
\\
\mathcal G_4&=F_7^{(1;2,0)}=\sigma^{-6}\varepsilon_{abcd}R_{ef}{}^{cd}\phi^m\phi^n\phi^e\phi^a\phi_m{}^f\phi_n{}^b,\label{Gbasis-4}
\\
\mathcal G_5&=F_8^{(1;2,0)}=\sigma^{-6}\varepsilon_{abcd}R_{ef}{}^{cm}\phi^a\phi^e\phi_m\phi^n\phi_n{}^b\phi^{df},\label{Gbasis-5}
\\
\mathcal G_6&=E_2^{(0;1,0)}F_1^{(1;1,0)}
=\sigma^{-6}\varepsilon_{abcd}R_{ef}{}^{cd}\phi^a\phi^e\phi^{bf}\phi_m\phi_n\phi^{mn},\label{Gbasis-6}
\\
\mathcal G_7&=F_4^{(2;0,0)}=\sigma^{-2}\varepsilon_{abcd}R_{ef}{}^{cd}R^{ae}\phi^b\phi^f.\label{Gbasis-7}
\end{align}
We refer to Ref.~\cite{Gao:2020juc} for complete lists of monomials and their expressions.
Here $E_1^{(0;1,0)}=\square\phi/\sigma$ and
$E_2^{(0;1,0)}=\phi^a\phi^b\phi_{ab}/\sigma^3$. 
We also select another three monomials:
\begin{align}
\mathcal G_8&=F_6^{(0;2,1)}=\sigma^{-6}\varepsilon_{abcd}\phi^e\phi_f\phi^a
\phi_e{}^b\phi_m{}^c\nabla^m\phi^{df},\label{GSTextra3-1}
\\
\mathcal G_9&=F_4^{(1;0,1)}=\sigma^{-4}\varepsilon_{abcd}R_{ef}{}^{cd}
\phi^a\phi^e\phi^m\nabla^b\phi_m{}^f,\label{GSTextra3-2}
\\
\mathcal G_{10}&=F_5^{(2;0,0)}=\sigma^{-4}\varepsilon_{abcd}R_{ef}{}^{cd}R^{amen}
\phi^b\phi^f\phi_m\phi_n.\label{GSTextra3-3}
\end{align}
As we will see in Eqs.~\eqref{GSTweighted-1}--\eqref{GSTweighted-3}, 
these three monomials are algebraically independent of $\mathcal{G}_{1},\cdots, \mathcal{G}_{7}$, but can be reduced by integrations by parts. 

Ref.~\cite{Gao:2020juc} exhausted the PV GST monomials at $d=4$ and identified fifteen unfactorizable and two factorizable and thus in total seventeen monomials. The remaining seven monomials are redundant by the identities below, with their labels inherited from Ref.~\cite{Gao:2020juc}. 
One can verify that these seventeen monomials satisfy the following seven algebraic identities:
\begin{align}
F_3^{(1;2,0)}&=\mathcal G_4-2\mathcal G_5-\mathcal G_6,\label{sevenidentities-1}
\\
F_4^{(1;2,0)}&=2\mathcal G_3-\mathcal G_4+2\mathcal G_5+\mathcal G_6,\label{sevenidentities-2}
\\
F_1^{(1;2,0)}&=4\mathcal G_2+4\mathcal G_3-2\mathcal G_4+4\mathcal G_5+2\mathcal G_6,\label{sevenidentities-3}
\\
E_1^{(0;1,0)}F_1^{(1;1,0)}&=\mathcal G_1+2\mathcal G_2+\mathcal G_4-2\mathcal G_5-\mathcal G_6,\label{sevenidentities-4}
\\
F_2^{(2;0,0)}&=4\mathcal G_{10},\label{sevenidentities-5}
\\
F_1^{(2;0,0)}&=8\mathcal G_7+16\mathcal G_{10},\label{sevenidentities-6}
\\
F_3^{(2;0,0)}&=-2\mathcal G_7-4\mathcal G_{10}.\label{sevenidentities-7}
\end{align}
As a result, the above ten monomials 
\begin{equation}
	\left\{\mathcal G_1,\ldots,\mathcal G_{10}\right\}
	\label{algebra10}
\end{equation}
are algebraically independent.
In Appendix~\ref{app:algebra}, we also prove completeness and linear independence of these ten monomials by an invertible component decomposition and independent variations of the local scalar and metric derivatives. 

The preceding monomials contain only $d=1$ and $d=2$ building blocks. The grading in Eq.~\eqref{grading} also allows the partitions $3+1$ and $4+0$. The $d=3$ blocks are $\nabla R$ and $\nabla^4\phi$, whereas the $d=4$ blocks are $\nabla^2R$ and $\nabla^5\phi$. The $3+1$ partition supplies three additional independent monomials:
\begin{align}
\mathcal J_1&=\sigma^{-2}\varepsilon_{abcd}\phi^a\phi^{be}\nabla^cR_e{}^d,\label{curvaturederivatives-1}
\\
\mathcal J_2&=\sigma^{-4}\varepsilon_{abcd}\phi^a\phi^e\phi^f\phi^{bm}\nabla^cR_{emf}{}^d,\label{curvaturederivatives-2}
\\
\mathcal J_3&=\sigma^{-4}\varepsilon_{abcd}\phi^a\phi^m\phi_m{}^b\phi^e\nabla^fR_{ef}{}^{cd}.\label{curvaturederivatives-3}
\end{align}
All three monomials contain one first derivative of curvature and one second derivative of the scalar field. In the full notation of Ref.~\cite{Gao:2020juc}, their class is $(c_0,c_1,c_2;d_2,d_3,d_4)=(0,1,0;1,0,0)$. 
This class of monomials is not considered in Ref.~\cite{Gao:2020juc} as all these three monomials can be reduced by integrations by parts\footnote{Ref.~\cite{Gao:2020juc} lists this class as reducible by integrations by parts (see Table~I), which is thus not taken into account in the subsequent analysis.}.
However, from the point of view of algebraic independence, no linear combination of $\mathcal G_1,\ldots,\mathcal G_{10}$ can reproduce these three monomials. This is briefly discussed in Appendix~\ref{app:algebra}, in which we show that \eqref{curvaturederivatives-1}-\eqref{curvaturederivatives-3} exhaust the independent contractions involving $d=3$ and $d=4$ building blocks.

To summarize, we follow Ref.~\cite{Gao:2020qxy} by denoting the linear spaces of all grade-$d$ PV GST polynomials by $\varXi_d$. In the sense of algebraic independence, we thus have\footnote{Eq.~\eqref{pointwisecount} thus also updates the results in Ref.~\cite{Gao:2020qxy} (see Eqs. (2.44) and (2.50)).}
\begin{equation}
\dim \varXi_3=1,\qquad \dim \varXi_4=13,
\label{pointwisecount}
\end{equation}
with $\varXi_3 = \mathrm{span}\{\mathcal G_0\}$ and 
\begin{equation}
	\varXi_4 = \mathrm{span} \{\mathcal G_1,\ldots,\mathcal G_{10},\mathcal J_1,\mathcal J_2,\mathcal J_3\}, \label{Xi4def}
\end{equation}
respectively.

\subsubsection{Integration-by-parts reduction}

We now examine integration-by-parts relations among the above monomials with arbitrary coefficients $f(\phi,X)$. Here the boundary terms are divergences of local tensor currents, with coefficients included inside the derivatives. This is necessary to determine the linearly independent monomials that are needed to construct a general polynomial-type Lagrangian. 

One can check that $\mathcal{G}_{8}$, $\mathcal{G}_9$ and $\mathcal{G}_{10}$ can be reduced by
\begin{align}
f\mathcal G_8&=-\tfrac12f\mathcal G_4+\nabla_a(fj_r^a),\label{GSTweighted-1}
\\
f\mathcal G_9&=-f\mathcal G_1-2f\mathcal G_3+(-3f+2Xf_X)\mathcal G_4
-2f\mathcal G_5-f\mathcal G_6+\nabla_a(fJ_s^a),\label{GSTweighted-2}
\\
f\mathcal G_{10}&=-2f\mathcal G_2-2f\mathcal G_3
+Xf_X\mathcal G_4-2Xf_X\mathcal G_5-Xf_X\mathcal G_6
-\tfrac12\sigma f_\phi\mathcal G_0\notag\\
&\quad+\nabla_a(fJ_D^a-2Xf_Xj_r^a),\label{GSTweighted-3}
\end{align}
where we have deliberately kept the total derivative terms, with currents given by
\begin{align}
j_r^a&=\sigma^{-6}\varepsilon_{bcd}{}^a\phi^e\phi^q\phi^b
\phi_e{}^c\phi^{dm}\phi_{mq},\label{GSTcurrents-1}
\\
J_s^a&=\sigma^{-4}g^{ab}\varepsilon_{cbde}R_{mn}{}^{de}
\phi^c\phi^m\phi^q\phi_q{}^n,\label{GSTcurrents-2}
\\
J_D^a&=2u^e\varepsilon_{ebc}{}^a(\mathrm{D}^ba^d+a^ba^d)K_d{}^c
+u^au_m\varepsilon^{mbcd}K_b{}^e\mathrm{D}_cK_{de}
+u_e\varepsilon^{eabc}K_b{}^dL_{cd}.\label{GSTcurrents-3}
\end{align}
In the above, the extrinsic curvature $K_{ab}$ and acceleration $a_{a}$ are defined with respect to the scalar foliation. That is, with normal $u_a=-\phi_a/\sigma$ and induced metric $h_{ab}=g_{ab}+u_au_b$, we use
\begin{equation}
	K_{ab}=-\sigma^{-1}h_a{}^ch_b{}^d\phi_{cd},\quad 
	a_a=\sigma^{-2}h_a{}^b\phi^c\phi_{cb}.
\end{equation}
Clearly, Eqs.~\eqref{GSTweighted-1}--\eqref{GSTweighted-3} imply that at the level of the Lagrangian, $\mathcal{G}_{8}$, $\mathcal{G}_{9}$ and $\mathcal{G}_{10}$ are not independent and can be reduced in terms of linear combinations of  $\mathcal G_0,\ldots,\mathcal G_7$ up to total derivatives\footnote{The coefficient derivatives follow from $(\nabla_af)j_r^a=0$ and $(\nabla_af)J_s^a=-2Xf_X\mathcal G_4$. The first two relations reproduce the reductions in Appendix~C of Ref.~\cite{Hu:2024hzo}, including their boundary terms. Defining $\mathcal D=\nabla_aJ_D^a=\mathcal G_{10}+2\mathcal G_2+2\mathcal G_3$ as in Appendix~B of Ref.~\cite{Hu:2024hzo}, the right-hand sides of Eqs.~(C28) and (C29) of Ref.~\cite{Hu:2024hzo} are $+16\mathcal D$ and $+4\mathcal D$, respectively. Subtracting four times the latter relation from the former gives an algebraic identity. For general $f$, the reduction of $f\mathcal D$ also contains the coefficient derivatives in Eq.~\eqref{GSTweighted-3}.}. Note that in \eqref{GSTweighted-3}, there is a term proportional to the $d=3$ monomial $\mathcal{G}_{0}$, which is due to the $\phi$-dependence of the coefficient function.

Moreover, the three monomials involving $d=3$ building blocks in Eqs.~\eqref{curvaturederivatives-1}--\eqref{curvaturederivatives-3} can also be reduced explicitly. By using
\begin{equation}
\nabla_a(f\sigma^{-p})=\sigma^{-p}\left[f_\phi\phi_a+
\left(f_X-\frac{p f}{2X}\right)\nabla_aX\right],
\qquad \nabla_aX=-\phi^b\phi_{ab},
\label{weightedpower}
\end{equation}
together with the commutator identity and Eqs.~\eqref{GSTweighted-1}--\eqref{GSTweighted-3}, one finds
\begin{align}
f\mathcal J_1&=(2f-2Xf_X)\mathcal G_3-\frac f2\mathcal G_7
+\nabla^c\left(f\sigma^{-2}\varepsilon_{abcd}\phi^a\phi^{be}R_e{}^d\right),\label{curvatureIBP-1}
\\
f\mathcal J_2&=-2f\mathcal G_2+f\mathcal G_3-\frac{Xf_X}{2}\mathcal G_4
+(4f-Xf_X)\mathcal G_5+\frac{Xf_X}{2}\mathcal G_6
+\frac{\sigma f_\phi}{4}\mathcal G_0\notag\\
&\quad+\nabla_c\left(f\sigma^{-4}\varepsilon_{ab}{}^c{}_d
\phi^a\phi^e\phi^f\phi^{bm}R_{emf}{}^d
-\frac f2J_D^c+Xf_Xj_r^c\right),\label{curvatureIBP-2}
\\
f\mathcal J_3&=2f\mathcal G_3
+\nabla_f\left(f\sigma^{-4}\varepsilon_{abcd}\phi^a\phi^m\phi_m{}^b\phi^e R_e{}^{fcd}
-fJ_s^f\right).\label{curvatureIBP-3}
\end{align}
Here we also keep the explicit expressions for the total derivative terms, in which currents $j^c_r$, $J^f_s$ and $J^c_D$ are given in Eqs. \eqref{GSTcurrents-1}-\eqref{GSTcurrents-3}. 
All three equalities include the full currents and hold for arbitrary $f(\phi,X)$. 
Note $\mathcal{J}_{2}$ also induces the $d=3$ monomial $\mathcal{G}_{0}$ when the coefficient depends on $\phi$. As a result, at the level of the Lagrangian, $\mathcal{J}_{1}$, $\mathcal{J}_{2}$ and $\mathcal{J}_{3}$ can be completely reduced to $\mathcal G_0,\ldots,\mathcal G_7$, and are thus not independent.

To prove independence of the seven $d=4$ monomials $\mathcal{G}_{1},\cdots, \mathcal{G}_{7}$ under integrations by parts, we must also exclude a tensor divergence relating the retained monomials. The classification of currents  in Appendix~\ref{app:independence}, together with the invertible GST-SCG correspondence discussed below, excludes such a relation (even when it contains a term proportional to $\mathcal G_0$). 

To summarize, up to $d=4$, the most general polynomial-type Lagrangian  of the parity-violating scalar-tensor theory  takes the form
\begin{equation}
S_{\rm PV}=\int\mathrm{d}^4x\sqrt{-g}\,\mathcal L_{\rm PV},
\label{GSTaction}
\end{equation}
where the Lagrangian is
\begin{equation}
\mathcal L_{\rm PV}=b_0(\phi,X)\mathcal G_0+\sum_{A=1}^7b_A(\phi,X)\mathcal G_A,
\label{GSTactionbasis}
\end{equation}
with $\mathcal{G}_{0}$ defined in \eqref{Gzero}, and $\mathcal{G}_{1},\cdots, \mathcal{G}_{7}$ defined in Eqs.~\eqref{Gbasis-1}-\eqref{Gbasis-7}.
Eq.~\eqref{GSTactionbasis} provides a complete and independent basis under both linear-algebraic identities and  integration-by-parts relations. There are one $d=3$ monomial and seven $d=4$ monomials. Note the former must be retained because derivatives of general coefficients generate terms such as $-\sigma f_\phi\mathcal G_0/2$ in Eq.~\eqref{GSTweighted-3}.

\subsection{Complete basis of parity-violating SCG polynomials and its GST correspondence}
\label{subsec:SCGbasis}

\subsubsection{Linear space of SCG polynomials}

In the SCG framework, we write the metric as
\begin{equation}
	\mathrm{d}s^2=-N^2\mathrm{d}t^2+h_{ij}(\mathrm{d}x^i+N^i\mathrm{d}t)(\mathrm{d}x^j+N^j\mathrm{d}t),\qquad \sigma=N^{-1},
\end{equation}
and define 
\begin{align}
K_{ij}&=\frac{\dot h_{ij}-\mathscr L_{\vec N}h_{ij}}{2N},\qquad
a_i=\mathrm{D}_i\ln N,\qquad F=\frac{\dot N-N^i\mathrm{D}_iN}{N^2},\label{ADMdefinitions-1}
\\
L_{ij}&\equiv h_i{}^ah_j{}^b\mathscr L_u K_{ab}
=\frac{\dot K_{ij}-\mathscr L_{\vec N}K_{ij}}{N}.\label{ADMdefinitions-2}
\end{align}
Here ${}^3R_{ij}$ denotes the intrinsic Ricci tensor and $K=h^{ij}K_{ij}$. The derivative $\mathrm{D}_i$ is the Levi--Civita derivative of $h_{ij}$, $\epsilon_{ijk}=u^a\varepsilon_{aijk}$, and spatial indices are raised with $h^{ij}$. A Lie derivative acts on both lower indices of $K_{ij}$. The covariant forms of these blocks are
\begin{align}
	F&=-\sigma^{-3}\phi^a\phi^b\phi_{ab},\notag\\
	L_{ab}&=h_a{}^ch_b{}^du^e\nabla_eK_{cd}+2K_a{}^cK_{cb}.
	\label{covariantblocks}
\end{align}
For a spatial tensor $S_{bc}$, $\mathrm{D}_a S_{bc}=h_a{}^dh_b{}^eh_c{}^f\nabla_dS_{ef}$. The derivative $\mathrm D_a$ extends to other ranks by projecting every index.

In the SCG framework, we also use the integer $d$ to classify SCG building blocks and monomials. $K_{ij},a_i,F$ are of $d=1$, and $\mathrm{D}_i$ or $\mathscr L_u$ raises $d$ by one.
Thus,  there are three types of $d=1$ building blocks: $K_{ij},a_i,F$. There are six types of $d=2$ building blocks: $L_{ij},{}^3R_{ij},\mathrm{D}_ia_j,\mathrm{D}_iK_{jk},\mathrm{D}_iF$, and $\mathscr L_uF$. The relation $\mathscr L_ua_i=\mathrm{D}_iF+Fa_i$ removes a redundant block. Also, $\mathrm{D}_ia_j$ is symmetric and the intrinsic Riemann tensor is determined by ${}^3R_{ij}$ in three dimensions. 
Building blocks of $d=3$ and $d=4$ are obtained by further spatial and projected Lie derivatives. 

As in the discussion of GST monomials, we first enumerate contractions of these blocks in the sense of linear-algebraic independence.

There is no PV scalar below $d=3$. The unique $d=3$ contraction is
\begin{equation}
\chi=\epsilon^{ijk}K_i{}^l\mathrm{D}_jK_{kl}.
\label{chidef}
\end{equation}
At $d=4$, we choose the following seven PV SCG monomials:
\begin{align}
\mathcal O_1&=\epsilon_{ijk}\,{}^3R^i{}_lK^{jl}a^k,\label{O1def}\\
\mathcal O_2&=\epsilon_{ijk}K^{im}K^{jn}\mathrm{D}_mK^k{}_n,\label{O2def}\\
\mathcal O_3&=\epsilon_{ijk}K^{mn}K^i{}_m\mathrm{D}^jK^k{}_n,\label{O3def}\\
\mathcal O_4&=\epsilon_{ijk}K^i{}_la^j\mathrm{D}^ka^l,\label{O4def}\\
\mathcal O_5&=\epsilon_{ijk}\,{}^3R^i{}_l\mathrm{D}^jK^{kl},\label{O5def}\\
\mathcal O_6&=\epsilon_{ijk}a^iK^{jl}L^k{}_l,\label{O6def}\\
\mathcal O_7&=F\epsilon_{ijk}K^i{}_l\mathrm{D}^jK^{kl}.\label{O7def}
\end{align}
For later convenience, we also define three additional $d=4$ contractions:
\begin{align}
\mathcal O_8&=\epsilon^{ijk}a_i a^l\mathrm{D}_kK_{lj},\notag\\
\mathcal O_9&=\epsilon^{ijk}(\mathrm{D}_i a^l)\mathrm{D}_kK_{lj},\notag\\
\mathcal O_{10}&=\epsilon^{ijk}L_i{}^l\mathrm{D}_jK_{kl},
\label{SCGextra3}
\end{align}
which are independent in the sense of algebraic identities but can be reduced by integrations by parts.

The single $d=3$ monomial $\chi$ and the ten $d=4$ monomials $\mathcal O_1,\ldots,\mathcal O_{10}$ exhaust the PV scalar contractions of the above building blocks of $d=1$ and $d=2$, as can be seen by the following enumeration. At $d=4$, their possible combinations have partitions $1+1+1+1$, $2+1+1$, and $2+2$, where each integer stands for the value of $d$ of one factor. 
These are also the three types used in the construction in Ref.~\cite{Hu:2024hzo}. In the first partition, all contractions vanish due to the antisymmetry of the Levi-Civita tensor. In the second partition, pairing $Ka$ with $L_{ij},{}^3R_{ij},\mathrm{D}_ia_j$ yields $\mathcal O_6,\mathcal O_1,\mathcal O_4$. Pairing $\mathrm{D}K$ with $aa$ or $KF$ yields $\mathcal O_8$ or $\mathcal O_7$. Pairing  $\mathrm{D}K$ with $KK$ yields the three contractions
$\mathcal O_2$, $\mathcal O_3$ and $K\chi$.
Finally, the $2+2$ partition gives $\mathcal O_{10},\mathcal O_5,\mathcal O_9$. Terms containing $\mathrm{D}_iF$ or $\mathscr L_uF$ vanish by index symmetry.

At this point, note that Ref.~\cite{Hu:2024hzo} includes all three $KK\mathrm{D}K$ contractions (i.e., $\mathcal O_2$, $\mathcal O_3$ and $K\chi$) in its $d=4$ list (see Eq. (24)). However, as pointed out in Ref.~\cite{Song:2026ajt}, only two of them are linearly independent. This can be seen from the fact that, for any $K^i{}_j$, the volume form satisfies
\begin{equation}
K_i{}^p\epsilon_{pjk}+K_j{}^p\epsilon_{ipk}+K_k{}^p\epsilon_{ijp}
=K\epsilon_{ijk}.
\label{volumeKidentity}
\end{equation}
Contracting this identity with $K^{il}\mathrm{D}^jK^k{}_l$ yields the identity
\begin{equation}
K\chi=\mathcal O_3-\mathcal O_2.
\label{cubicidentity}
\end{equation}
Thus $K\chi$ is not independent.

Similar to the discussion of GST monomials, the $3+1$ partition of $d=4$ must also be included. Commuting spatial and projected Lie derivatives, using $\mathscr L_ua_i=\mathrm D_iF+Fa_i$, and expressing $\mathscr L_u{}^3R_{ij}$ through spatial derivatives of $NK_{ij}$ leaves the following $d=3$ building blocks
\begin{equation}
\begin{gathered}
\mathrm D_iL_{jk},\quad \mathscr L_uL_{ij},\quad \mathrm D_i{}^3R_{jk},\quad
\mathrm D_i\mathrm D_j a_k,\quad \mathrm D_i\mathrm D_jK_{kl},\\
\mathrm D_i\mathrm D_jF,\quad \mathrm D_i\mathscr L_uF,\quad \mathscr L_u^2F,
\end{gathered}
\label{SCGthirdblocks}
\end{equation}
together with products of lower-grade blocks. Pairing $\mathrm D L$ or $\mathrm D{}^3R$ with $K$ gives one PV contraction each. Pairing $\mathrm D\mathrm D K$ with $a$ gives one further direction, up to curvature commutators. The $(\mathrm D\mathrm D a)K$ contractions reduce to ${}^3RKa$, and the remaining pairings vanish by index symmetry. 
Finally, we find  three nonvanishing and independent contractions: 
	\begin{align}
		\mathcal P_1&=\epsilon^{ijk}K_i{}^l\mathrm D_jL_{kl},\label{P1}\\
		\mathcal P_2&=\epsilon^{ijk}K_i{}^l\mathrm D_j{}^3R_{kl},\label{P2}\\
		\mathcal P_3&=\epsilon^{ijk}a_i\mathrm D_j\mathrm D^lK_{kl}.
		\label{P3}
	\end{align}
We refer to Appendix~\ref{app:algebra} for a brief discussion on their linear independence.
The $4+0$ partition adds no independent contraction, although individual contractions need not vanish. For example,
\begin{equation}
	\epsilon^{ijk}\mathrm D_i\mathrm D^l\mathrm D_jK_{kl}
	=\mathcal P_2-\mathcal O_5.
	\label{SCGfourthblock}
\end{equation}
The reduction of the remaining single-block types is given in Appendix~\ref{app:algebra}.

To summarize, at $d=4$ we find thirteen linearly independent PV SCG monomials
\begin{equation}
\{\mathcal O_1,\ldots,\mathcal O_{10},\mathcal P_1,\mathcal P_2,\mathcal P_3\}.
\label{SCGalgebra13}
\end{equation}
Denoting the linear spaces of all grade-$d$ PV SCG polynomials  by $S_d$, we obtain
\begin{equation}
\dim S_3=1,\qquad \dim S_4=13.
\label{SCGpointwisecount}
\end{equation}
These counts agree with Eq.~\eqref{pointwisecount}.

For our purpose, we would like to examine the correspondence among polynomials (vectors) in both spaces. Here we show the explicit map from $\varXi_4$ to $S_4$. This can be done by using the decomposition of the second derivative of the scalar field as well as Gauss-Codazzi relations \cite{Gao:2020yzr,Gao:2020qxy,Hu:2021bbo}.
For $d=3$, we find 
	\begin{equation}
		\mathcal G_0=2\chi.
	\end{equation}
For $d=4$, we find
\begin{align}
\mathcal G_1&=-2(\mathcal O_8+\mathcal O_3),\notag\\
\mathcal G_2&=\mathcal O_2-\mathcal O_4-\mathcal O_6,\notag\\
\mathcal G_3&=-\mathcal O_1+\mathcal O_4+\mathcal O_6,\notag\\
\mathcal G_4&=2\mathcal O_8,\notag\\
\mathcal G_5&=-\mathcal O_4-\mathcal O_6,\notag\\
\mathcal G_6&=-2\mathcal O_7,\notag\\
\mathcal G_7&=2(\mathcal O_2-\mathcal O_5-\mathcal O_9-\mathcal O_8+\mathcal O_3-\mathcal O_{10}),\notag\\
\mathcal G_8&=-\mathcal O_4,\notag\\
\mathcal G_9&=2(\mathcal O_9-\mathcal O_8+\mathcal O_7+\mathcal O_3),\notag\\
\mathcal G_{10}&=2(\mathcal O_9+\mathcal O_8-\mathcal O_3+\mathcal O_{10}),
\label{rawdictionary}
\end{align}
and
\begin{align}
\mathcal J_1&=\mathcal P_1+\mathcal P_2-\mathcal P_3
-\mathcal O_2-\mathcal O_3+2\mathcal O_4+\mathcal O_6,\label{rawJ1}\\
\mathcal J_2&=-\mathcal P_1-\mathcal O_1-3\mathcal O_2
+\mathcal O_3-\mathcal O_4,\label{rawJ2}\\
\mathcal J_3&=-2\mathcal P_3-2\mathcal O_1+2\mathcal O_4+2\mathcal O_6.
\label{rawJ3}
\end{align}
Since the above map $\varXi_{4} \rightarrow S_4$ is invertible, it is straightforward to derive the inverse map.

We emphasize that the above relations are algebraic identities. Each can be multiplied by an independent arbitrary coefficient without generating coefficient derivatives. The two coefficient functions denoted by $f$ are related by
\begin{equation}
f_{\rm SCG}(t,N)=f_{\rm GST}\left(\phi,X\right),\qquad
Nf_{,N}=-2Xf_{,X},\qquad \frac{f_{,t}}{N}=\sigma f_{,\phi}.
\label{coefficientconventions}
\end{equation}

\subsubsection{Integration-by-parts reduction}

At the level of the action, the thirteen GST monomials admit the six integration-by-parts reductions in Eqs.~\eqref{GSTweighted-1}--\eqref{GSTweighted-3} and \eqref{curvatureIBP-1}--\eqref{curvatureIBP-3}. We therefore expect similar reductions in the SCG framework.

After some manipulations, we find\footnote{These relations reproduce the reductions of $aa\mathrm{D}K$, $(\mathrm{D}a)\mathrm{D}K$ and $L\mathrm{D}K$ discussed in Ref.~\cite{Hu:2024hzo}.}
\begin{align}
	f\mathcal O_8&=f\mathcal O_4+\nabla_a(fj_r^a),\label{weightedIBP-1}
	\\
	f\mathcal O_9&=f\mathcal O_1-(f+Nf_N)\mathcal O_4+\nabla_a(fj_s^a),\label{weightedIBP-2}
	\\
	f\mathcal O_{10}&=f(\mathcal O_3-\mathcal O_2)-\frac{Nf_N}{2}(\mathcal O_6+\mathcal O_7)
	-\frac{f_t}{2N}\chi+\frac12\nabla_a(fj_t^a).\label{weightedIBP-3}
\end{align}
Note that \eqref{weightedIBP-3} involves the $d=3$ monomial $\chi$, if the coefficient $f$ depends on $t$ explicitly.
In the above, the currents are given by
\begin{align}
j_r^a&=u^e\varepsilon_{ebc}{}^a a^b a^dK_d{}^c,\notag\\
j_s^a&=u^e\varepsilon_{ebc}{}^a \mathrm{D}^ba^dK_d{}^c,\notag\\
j_t^a&=u^a\chi+u_e\varepsilon^{eabc}K_b{}^dL_{cd}.
\label{currents}
\end{align}
The expression for $j_r^a$ agrees with Eq.~\eqref{GSTcurrents-1}, and $J_D^a$ in Eq.~\eqref{GSTcurrents-3} obeys
\begin{equation}
	J_D^a=2j_s^a+2j_r^a+j_t^a.
	\label{JDspatial}
\end{equation}

For the three monomials involving $d=3$ building blocks, one finds
\begin{align}
f\mathcal P_1&=f(\mathcal O_3-\mathcal O_2)
+\left(f+\frac{Nf_N}{2}\right)\mathcal O_6
-\frac{Nf_N}{2}\mathcal O_7-\frac{f_t}{2N}\chi
+\nabla_a(f\mathcal C_1^a),\label{higherIBP1}\\
f\mathcal P_2&=f\mathcal O_5-(f+Nf_N)\mathcal O_1
+\nabla_a(f\mathcal C_2^a),\label{higherIBP2}\\
f\mathcal P_3&=\nabla_a(f\mathcal C_3^a),\label{higherIBP3}
\end{align}
where the currents are
\begin{align}
\mathcal C_1^a&=\frac12\left(u^a\chi-u_e\varepsilon^{eabc}K_b{}^dL_{cd}\right),\label{highercurrent1}\\
\mathcal C_2^a&=-u_e\varepsilon^{eabc}K_b{}^d{}^3R_{cd},\label{highercurrent2}\\
\mathcal C_3^a&=-u_e\varepsilon^{eabc}a_b\mathrm D^dK_{cd}.\label{highercurrent3}
\end{align}
All spatial tensors in these currents are fully projected. 

Eqs.~\eqref{weightedIBP-1}--\eqref{weightedIBP-3} and Eqs.~\eqref{higherIBP1}--\eqref{higherIBP3} imply that $\mathcal O_8,\mathcal O_9,\mathcal O_{10}$ and $\mathcal{P}_1,\mathcal{P}_2,\mathcal{P}_3$ are not independent at the level of the Lagrangian.
In Appendix~\ref{app:independence}, we briefly show that no additional integration-by-parts relations relate the retained monomials.
As a result, up to $d=4$, the most general polynomial-type Lagrangian  of the parity-violating SCG theory  takes the form
\begin{equation}
S_{\rm PV}=\int \mathrm{d}t\,\mathrm{d}^3x\,N\sqrt h
\left[c_0(t,N)\chi+\sum_{A=1}^7c_A(t,N)\mathcal O_A\right].
\label{PVactionbasis}
\end{equation}
Eq.~\eqref{PVactionbasis} can be compared with the GST action \eqref{GSTactionbasis}.

We are now ready to map the two general actions.
For a GST Lagrangian in Eq.~\eqref{GSTactionbasis}, by combining the mappings \eqref{rawdictionary} and the integration-by-parts relations Eqs.~\eqref{weightedIBP-1}--\eqref{weightedIBP-3} and Eqs.~\eqref{higherIBP1}--\eqref{higherIBP3}, we find
\begin{equation}
\sum_{A=0}^7 b_A\mathcal G_A
=c_0\chi+\sum_{A=1}^7c_A\mathcal O_A
+\nabla_a\left[2(b_4-b_1)j_r^a-b_7J_D^a\right].
\label{fullactiondictionary-1}
\end{equation}
The maps between the two sets of coefficients are
\begin{align}
c_0&=2b_0+\sigma b_{7,\phi},\notag\\
c_1&=-b_3-2b_7,\notag\\
c_2&=b_2+4b_7,\notag\\
c_3&=-2b_1,\notag\\
c_4&=-2b_1-b_2+b_3+2b_4-b_5-4Xb_{7,X},\notag\\
c_5&=-2b_7,\notag\\
c_6&=-b_2+b_3-b_5-2Xb_{7,X},\notag\\
c_7&=-2b_6-2Xb_{7,X},
\label{forwardcoeff}
\end{align}
and 
\begin{align}
b_0&=c_0/2+\sigma c_{5,\phi}/4,\notag\\
b_1&=-c_3/2,\notag\\
b_2&=c_2+2c_5,\notag\\
b_3&=-c_1+c_5,\notag\\
b_4&=(-c_3+c_4-c_6-Xc_{5,X})/2,\notag\\
b_5&=-c_1-c_2-c_5-c_6+Xc_{5,X},\notag\\
b_6&=(-c_7+Xc_{5,X})/2,\notag\\
b_7&=-c_5/2.
\label{inversecoeff}
\end{align}
Note that in \eqref{fullactiondictionary-1} we also deliberately keep the total derivative term.

\subsection{The subsets without higher time derivatives}
The correspondence established above allows us to select the GST combinations that contain no higher time derivatives in unitary gauge by first identifying the corresponding SCG subset. At $d=3$, both $\chi$ and $\mathcal G_0=2\chi$ already satisfy this requirement. At $d=4$, we select the combinations of the seven SCG monomials that contain neither the lapse velocity $F$ nor the projected Lie derivative $L_{ij}$ of the extrinsic curvature, and then recover their covariant expressions.

The resulting SCG subset without higher time derivatives is
\begin{equation}
\{\chi,\mathcal O_1,\mathcal O_2,\mathcal O_3,\mathcal O_4,\mathcal O_5\}.
\label{safeSCG}
\end{equation}
The definitions in Eqs.~\eqref{chidef} and \eqref{O1def}--\eqref{O5def} show explicitly that these monomials contain neither $F$ nor $L_{ij}$. The corresponding GST subset consists of $\mathcal G_0$ and five $d=4$ combinations. Retaining the labels of Ref.~\cite{Hu:2024hzo}, we choose\footnote{The label $\mathcal L_1$ is reserved for the $d=3$ term $\mathcal G_0$ in Ref.~\cite{Hu:2024hzo}.  $\mathcal L_6$ in Eq.~(57) of Ref.~\cite{Hu:2024hzo} obeys the algebraic identity $\mathcal L_6=\mathcal L_2+\mathcal L_3+2\mathcal L_5=-2K\chi$ and is thus not independent.} 
\begin{align}
\mathcal L_2&=\mathcal G_1
=\sigma^{-4}\varepsilon_{abcd}R_{ef}{}^{cd}\phi^a\phi^e\phi_m{}^b\phi^{fm},\label{safeL2}\\
\mathcal L_3&=\mathcal G_4
=\sigma^{-6}\varepsilon_{abcd}R_{ef}{}^{cd}\phi^m\phi^n\phi^e\phi^a\phi_m{}^f\phi_n{}^b,\label{safeL3}\\
\mathcal L_4&=\mathcal G_3+\mathcal G_5\notag\\
&=\sigma^{-4}\varepsilon_{abcd}R^{amen}\phi^b\phi^f\phi_e{}^c\phi_f{}^d
\left(g_{mn}+\frac{\phi_m\phi_n}{\sigma^2}\right),\label{safeL4}\\
\mathcal L_5&=\mathcal G_2-\mathcal G_5\notag\\
&=\sigma^{-4}\varepsilon_{abcd}R_{ef}{}^{cm}\phi^a\phi^e\phi^{bf}\phi^{dn}
\left(g_{mn}+\frac{\phi_m\phi_n}{\sigma^2}\right),\label{safeL5}\\
\mathcal L_7&=\mathcal G_7+\mathcal G_{10}\notag\\
&=\sigma^{-2}\varepsilon_{abcd}R_{ef}{}^{cd}R^{amen}\phi^b\phi^f
\left(g_{mn}+\frac{\phi_m\phi_n}{\sigma^2}\right).\label{safeL7}
\end{align}
Consequently, the subset without higher time derivatives contains one $d=3$ and five independent $d=4$ combinations.

It is also useful to derive the inverse correspondence.
For an arbitrary coefficient $f$, we find
\begin{align}
f\mathcal L_2&=-2f\epsilon_{ijk}\left(K^{mn}K^i{}_m\mathrm{D}^jK^k{}_n
+K^i{}_la^j\mathrm{D}^ka^l\right)-2\nabla_a(fj_r^a),\label{safeweighted2}\\
f\mathcal L_3&=2f\epsilon_{ijk}K^i{}_la^j\mathrm{D}^ka^l
+2\nabla_a(fj_r^a),\label{safeweighted3}\\
f\mathcal L_4&=-f\epsilon_{ijk}\,{}^3R^i{}_lK^{jl}a^k,\label{safeweighted4}\\
f\mathcal L_5&=f\epsilon_{ijk}K^{im}K^{jn}\mathrm{D}_mK^k{}_n,\label{safeweighted5}\\
f\mathcal L_7&=2f\epsilon_{ijk}\left(K^{im}K^{jn}\mathrm{D}_mK^k{}_n
-{}^3R^i{}_l\mathrm{D}^jK^{kl}\right).\label{safeweighted7}
\end{align}
Thus the five covariant combinations span precisely $\mathcal O_1,\ldots,\mathcal O_5$ at the action level.

The remaining two SCG action representatives, $\mathcal O_6$ and $\mathcal O_7$ in Eqs.~\eqref{O6def} and \eqref{O7def}, contain $L_{ij}$ and $F$, respectively. By Eqs.~\eqref{ADMdefinitions-1} and \eqref{ADMdefinitions-2}, these introduce a second time derivative of the spatial metric and a first time derivative of the lapse.
For an SCG action depending on $N,h_{ij},K_{ij}$ and their spatial derivatives, with no $F$ or $L_{ij}$, such theories generically propagate two tensor modes and one scalar mode \cite{Gao:2014soa,Gao:2014fra}. 
Adding $\mathcal O_7$ can make the lapse dynamical, while adding $\mathcal O_6$ can require the extrinsic curvature to be treated as an independent coordinate. Although both interactions are linear in the additional velocities, the number of degrees of freedom in the presence of such terms must be examined. This will be the main task in the remaining part of this paper.

\section{The lapse-velocity interaction}
\label{sec:F}

We consider the following action
\begin{equation}
	S=\int \mathrm{d}t\,\mathrm{d}^3x\,N\sqrt h\left[{}^3R+K_{ij}K^{ij}-K^2+c_3
	+c_1F\chi+c_2\epsilon^{ijk}a_iK_j{}^l L_{kl}\right].
	\label{masteraction}
\end{equation}
Here $c_1,c_2,c_3$ are functions of $(t,N)$ and denote the coefficients of this action, independently of the basis coefficients in Eq.~\eqref{PVactionbasis}. With $c_1=c_2=0$, the theory has 3 DOFs as long as $\partial_N^2(Nc_3)\neq 0$.
Starting from \eqref{masteraction}, we will determine whether the two interactions, separately or together, add canonical modes or admit a degeneracy that can reduce the number of DOFs.

\subsection{Primary constraints and total Hamiltonian}

We first focus on the lapse-velocity term $F\chi$ by setting $c_2=0$ in Eq.~\eqref{masteraction}. Following Refs.~\cite{Gao:2018znj,Gao:2019twq}, we introduce auxiliary variables $A$ and $Q_{ij}$ by imposing $A=F$ and $Q_{ij}=K_{ij}$ with Lagrange multipliers $\lambda$ and $\lambda_{(1)}^{ij}$. The equivalent first-order Lagrangian is
\begin{align}
\widetilde L_F=\int \mathrm{d}^3x\bigg\{&N\sqrt h[{}^3R+Q_{ij}Q^{ij}-Q^2+c_3+c_1A\chi]
+\frac{\lambda_{(1)}^{ij}}2(\dot h_{ij}-\mathscr L_{\vec N}h_{ij})-N\lambda_{(1)}^{ij}Q_{ij}\notag\\
&+\lambda(\dot N-N^i\mathrm{D}_iN)-N^2\lambda A\bigg\},
\label{Fauxaction}
\end{align}
where $Q=h^{ij}Q_{ij}$ and from now on $\chi=\epsilon^{ijk}Q_i{}^l\mathrm{D}_jQ_{kl}$. The 24 configuration variables are $\varPhi^I=\{N,\allowbreak N^i,\allowbreak A,\allowbreak h_{ij},\allowbreak Q_{ij},\allowbreak \lambda,\allowbreak \lambda_{(1)}^{ij}\}$. We denote their conjugate momenta as $\varPi_I=\{p,\allowbreak p_i,\allowbreak \pi,\allowbreak p_{(1)}^{ij},\allowbreak p_{(2)}^{ij},\allowbreak \varPi,\allowbreak \varPi^{(1)}_{ij}\}$. 
The Poisson bracket is thus defined by
\begin{equation}
	[U,V]=\sum_I\int \mathrm{d}^3x\left(\frac{\delta U}{\delta \varPhi^I(\vec{x})}\frac{\delta V}{\delta \varPi_I(\vec{x})}
	-\frac{\delta U}{\delta \varPi_I(\vec{x})}\frac{\delta V}{\delta \varPhi^I(\vec{x})}\right).
	\label{PB}
\end{equation}

The definition $\varPi_I=\delta\widetilde L_F/\delta\dot\varPhi^I$ gives
\begin{equation}
p=\lambda,\quad p_i=0,\quad\pi=0,\quad
p_{(1)}^{ij}=\tfrac12\lambda_{(1)}^{ij},\quad
p_{(2)}^{ij}=0,\quad\varPi=0,\quad\varPi^{(1)}_{ij}=0.
\label{Fmomenta}
\end{equation}
Note that the momenta conjugate to ordinary tensors have density weight one, while $\varPi,\varPi^{(1)}_{ij}$ have weight zero. According to \eqref{Fmomenta}, all the momentum equations are primary constraints,
\begin{equation}
\begin{gathered}
\phi_N\equiv p-\lambda\approx0,\qquad p_i\approx0,\qquad\pi\approx0,\qquad
\phi_h^{ij}\equiv p_{(1)}^{ij}-\tfrac12\lambda_{(1)}^{ij}\approx0,\\
p_{(2)}^{ij}\approx0,\qquad\varPi\approx0,\qquad\varPi^{(1)}_{ij}\approx0.
\end{gathered}
\label{Fprimary}
\end{equation}

For later convenience, we define the following functionals:
\begin{equation}
\varphi[W]=\int\mathrm{d}^3x\,W\varphi,\qquad
\varphi[Z]=\int\mathrm{d}^3x\,Z_{ij}\varphi^{ij}
\label{smearingdefinition}
\end{equation}
where $\varphi$ is a scalar (or scalar density) and  $\varphi^{ij}$ is a tensor (or tensor density), and $W$ and $Z_{ij}$ are smearing functions.
The smearing functions are independent of the canonical variables and have the tensor type and density weight needed to make each integrand a scalar density.

Since our theory has spatial diffeomorphism invariance, it is convenient to split the canonical Hamiltonian into two parts.
We denote the part of the canonical Hamiltonian independent of the shift vector  by $H_F$, where the subscript identifies the interaction containing the lapse velocity $F$. It is
\begin{equation}
H_F=\int \mathrm{d}^3x\left[N\lambda_{(1)}^{ij}Q_{ij}+N^2\lambda A
-N\sqrt h({}^3R+Q_{ij}Q^{ij}-Q^2+c_3+c_1A\chi)\right].
\label{HF}
\end{equation}
For a field-independent spatial vector $\zeta^i$, the generator of spatial diffeomorphisms is
\begin{equation}
\mathcal C[\zeta]=\int \mathrm{d}^3x\sum_I\varPi_I\mathscr L_{\vec\zeta}\varPhi^I
=\int \mathrm{d}^3x\,\zeta^i\mathcal C_i,
\label{spatialgeneratorsmeared}
\end{equation}
where
\begin{align}
\mathcal C_i={}&p\mathrm{D}_iN+\pi \mathrm{D}_iA-2h_{ik}\mathrm{D}_jp_{(1)}^{jk}
+p_{(2)}^{jk}\mathrm{D}_iQ_{jk}-2\mathrm{D}_j(p_{(2)}^{jk}Q_{ki})\notag\\
&+p_k\mathrm{D}_iN^k+\mathrm{D}_j(p_iN^j)+\varPi \mathrm{D}_i\lambda-\mathrm{D}_i(\varPi\lambda)\notag\\
&+\varPi^{(1)}_{jk}\mathrm{D}_i\lambda_{(1)}^{jk}
+2\mathrm{D}_j(\varPi^{(1)}_{ik}\lambda_{(1)}^{jk})-\mathrm{D}_i(\varPi^{(1)}_{kl}\lambda_{(1)}^{kl}).
\label{spatialgenerator}
\end{align}
The total Hamiltonian is thus
\begin{align}
H_{\mathrm{T}}=H_F+\mathcal C[\vec N]+\int \mathrm{d}^3x\big(&\xi\phi_N+\xi^ip_i+\gamma\pi
+\nu^{(1)}_{ij}\phi_h^{ij}+\nu^{(2)}_{ij}p_{(2)}^{ij}
+\varXi\varPi+\mu^{ij}\varPi^{(1)}_{ij}\big).
\label{FHT}
\end{align}

\subsection{Primary preservation and secondary constraints}

A constraint $\varphi$ must be preserved in time evolution, which yields the so-called preservation equation given by $\dot\varphi=\partial_t\varphi+[\varphi,H_{\mathrm{T}}]\approx0$. Let $\varphi_A$ denote the primary constraints in Eq.~\eqref{Fprimary}, and let $\varLambda^A=\{\xi,\xi^i,\gamma,\nu^{(1)}_{ij},\nu^{(2)}_{ij},\varXi,\mu^{ij}\}$ denote their multipliers in Eq.~\eqref{FHT}, with $A$ running over all tensor components. Except for the momenta $p_i$ conjugate to the shift vector, this condition reads
\begin{equation}
	0\approx\dot\varphi(\vec{x})=\partial_t\varphi(\vec{x})+[\varphi(\vec{x}),H_F]
	+\mathscr L_{\vec N}\varphi(\vec{x})+\int \mathrm{d}^3y\,\varLambda^A(\vec{y})[\varphi(\vec{x}),\varphi_A(\vec{y})].
	\label{preservationrule}
\end{equation}
If the multiplier terms in a preservation equation determine the remaining multipliers, that equation generates no new constraint. A left null direction instead gives a compatibility condition on the source, which must be checked for a new independent constraint. 

Due to the last term in \eqref{preservationrule}, it is necessary to evaluate the Poisson brackets among all primary constraints. We denote the primary Poisson-bracket matrix as $\mathbb P_F(\vec{x},\vec{y})$, which is given by
\begin{equation}
\begingroup\setlength{\arraycolsep}{3pt}\renewcommand{\arraystretch}{1.6}
\begin{array}{c|ccccccc}
[\cdot,\cdot]&\phi_N(\vec{y})&p_j(\vec{y})&\pi(\vec{y})&\phi_h^{kl}(\vec{y})&p_{(2)}^{kl}(\vec{y})&\varPi(\vec{y})&\varPi^{(1)}_{kl}(\vec{y})\\\hline
\phi_N(\vec{x})&0&0&0&0&0&-\delta^3(\vec{x}-\vec{y})&0\\
p_i(\vec{x})&0&0&0&0&0&0&0\\
\pi(\vec{x})&0&0&0&0&0&0&0\\
\phi_h^{ij}(\vec{x})&0&0&0&0&0&0&-\tfrac12\delta^{ij}_{kl}\delta^3(\vec{x}-\vec{y})\\
p_{(2)}^{ij}(\vec{x})&0&0&0&0&0&0&0\\
\varPi(\vec{x})&\delta^3(\vec{x}-\vec{y})&0&0&0&0&0&0\\
\varPi^{(1)}_{ij}(\vec{x})&0&0&0&\tfrac12\delta^{kl}_{ij}\delta^3(\vec{x}-\vec{y})&0&0&0\\
\end{array}.
\endgroup
\label{Fprimarymatrix}
\end{equation}
Here $\delta^3(\vec{x}-\vec{y})$ is the spatial Dirac delta distribution and $\delta_{ij}^{kl}=\delta_i^{(k}\delta_j^{l)}$ is the symmetric tensor identity.

As a result, all 24 primary preservation conditions are
\begin{align}
\dot\phi_N&\approx-\frac{\delta H_F}{\delta N}-\varXi=0,&
\dot\phi_h^{ij}&\approx-\frac{\delta H_F}{\delta h_{ij}}-\frac12\mu^{ij}=0,\notag\\
\dot\varPi&\approx-N^2A+\xi=0,&
\dot\varPi^{(1)}_{ij}&\approx-NQ_{ij}+\frac12\nu^{(1)}_{ij}=0,\notag\\
\dot p_i&\approx-\mathcal C_i=0,&\dot\pi&\approx\alpha=0,&
\dot p_{(2)}^{ij}&\approx\beta^{ij}=0,
\label{Fprimarypreserve}
\end{align}
where
\begin{align}
\alpha&=-N^2\lambda+\sqrt hNc_1\chi,\label{alpha}\\
\beta^{ij}&=-N\lambda_{(1)}^{ij}+2N\sqrt h(Q^{ij}-Qh^{ij})\notag\\
&\quad+\sqrt h\epsilon^{(i|kl}\left[2Nc_1A \mathrm{D}_kQ_l{}^{|j)}+\mathrm{D}_k(Nc_1A)Q_l{}^{|j)}\right].
\label{beta}
\end{align}
Thus $\mathcal C_i\approx0$, $\alpha\approx0$, and $\beta^{ij}\approx0$ are ten secondary constraints, while the first four equations in Eq.~\eqref{Fprimarypreserve} fix 14 multipliers. This result is also consistent with the fact that the matrix $\mathbb P_F(\vec{x},\vec{y})$ has rank 14.  As in all SCG theories, the three constraints $\mathcal C_i\approx0$ generate spatial diffeomorphisms and are given explicitly in Eq.~\eqref{spatialgenerator}.

\subsection{Secondary preservation and number of degrees of freedom}

The secondary constraints $\alpha$, $\beta^{ij}$ and $\mathcal{C}_i$ must also be preserved in time.
In order to determine whether the preservation equations for the secondary constraints determine the Lagrange multipliers or yield new constraints, it is necessary to evaluate the Poisson brackets among the primary and secondary constraints.

First, note that $\alpha,\beta^{ij}$ and $H_F$ depend only on the configuration fields $\varPhi^I$ (together with the multipliers $\lambda,\lambda_{(1)}^{ij}$). They contain none of the conjugate momenta $\varPi_I$. Thus their mutual Poisson brackets and their brackets with $H_F$ vanish strongly.
Let $\boldsymbol{\varphi}_F=(\phi_N,p_i,\pi,\phi_h^{ij},p_{(2)}^{ij},\varPi,\varPi^{(1)}_{ij})$ denote the primary constraints in Eq.~\eqref{Fprimary}, and let $\boldsymbol{s}_F=(\alpha,\beta^{ij})$. The full Poisson bracket matrix of all the constraints is
\begin{equation}
	\begin{array}{c|ccc}
		[\cdot,\cdot]&\boldsymbol{\varphi}_F(\vec{y})&\boldsymbol{s}_F(\vec{y})&\mathcal C_j(\vec{y})\\\hline
		\boldsymbol{\varphi}_F(\vec{x})&\mathbb P_F&\mathbb B_F&0\\
		\boldsymbol{s}_F(\vec{x})&-\mathbb B_F^{\mathrm T}&0&0\\
		\mathcal C_i(\vec{x})&0&0&0
	\end{array}, \label{Ffullmatrix}
\end{equation}
where  $\mathbb P_F$ is the block of primary constraints given in Eq.~\eqref{Fprimarymatrix}, and
\begin{equation}
\mathbb B_F(\vec{x},\vec{y})=
\begin{array}{c|cc}
[\cdot,\cdot]&\alpha(\vec{y})&\beta^{kl}(\vec{y})\\\hline
\phi_N(\vec{x})&-\dfrac{\delta\alpha(\vec{y})}{\delta N(\vec{x})}&-\dfrac{\delta\beta^{kl}(\vec{y})}{\delta N(\vec{x})}\\
p_i(\vec{x})&0&0\\
\pi(\vec{x})&0&-\dfrac{\delta\beta^{kl}(\vec{y})}{\delta A(\vec{x})}\\
\phi_h^{ij}(\vec{x})&-\dfrac{\delta\alpha(\vec{y})}{\delta h_{ij}(\vec{x})}&-\dfrac{\delta\beta^{kl}(\vec{y})}{\delta h_{ij}(\vec{x})}\\
p_{(2)}^{ij}(\vec{x})&-\dfrac{\delta\alpha(\vec{y})}{\delta Q_{ij}(\vec{x})}&-\dfrac{\delta\beta^{kl}(\vec{y})}{\delta Q_{ij}(\vec{x})}\\
\varPi(\vec{x})&N^2(\vec{y})\delta^3(\vec{x}-\vec{y})&0\\
\varPi^{(1)}_{ij}(\vec{x})&0&N(\vec{y})\delta_{ij}^{kl}\delta^3(\vec{x}-\vec{y})
\end{array}.
\end{equation}
Here a block transpose exchanges both spatial arguments and tensor indices, e.g., $\mathbb B^{\mathrm T}(\vec{x},\vec{y})=\mathbb B(\vec{y},\vec{x})^{\mathrm T}$. 
The displayed zeros involving $\mathcal C_i$ are weak equalities. All other blocks are exact. The entries in $\mathbb B_F$ are evaluated in Eqs.~\eqref{Fmultiplieroperator}, \eqref{alphabrackets}, and \eqref{betabrackets}.

Using Eq.~\eqref{Fprimarypreserve}, the preservation equations for $\alpha$ and $\beta^{ij}$ are
\begin{align}
0&=\partial_t\alpha+\delta_N\alpha[N^2A]+\delta_h\alpha[2NQ]
-N^2\varXi+\delta_Q\alpha[\nu^{(2)}],\\
0&=\partial_t\beta^{ij}+\delta_N\beta^{ij}[N^2A]
+\delta_h\beta^{ij}[2NQ]-N\mu^{ij}
+\delta_A\beta^{ij}[\gamma]+\delta_Q\beta^{ij}[\nu^{(2)}].
\label{Fsecondarypreserve}
\end{align}
The 14 already fixed multipliers are
\begin{equation}
\xi=N^2A,\qquad \nu^{(1)}_{ij}=2NQ_{ij},\qquad
\varXi=-\frac{\delta H_F}{\delta N},\qquad
\mu^{ij}=-2\frac{\delta H_F}{\delta h_{ij}}.
\label{Ffixedmultipliers}
\end{equation}
The seven remaining nonspatial multipliers are $\gamma$ and the six components of $\nu^{(2)}_{ij}$. The three shift multipliers $\xi^i$ remain arbitrary.
The variations $\delta_N\alpha[N^2A]$, $\delta_h\alpha[2NQ]$, and their counterparts for $\beta^{ij}$ are defined by Eq.~\eqref{variationdefinition} with the indicated variations of the fields. We refer to Appendix~\ref{app:variations} for the detailed calculations, together with the functional derivatives $\delta H_F/\delta N$ and $\delta H_F/\delta h_{ij}$ that fix $\varXi$ and $\mu^{ij}$. The terms containing $\gamma$ and $\nu^{(2)}_{ij}$ are displayed below, with $W=\gamma$ and $Z_{ij}=\nu^{(2)}_{ij}$:
\begin{align}
\delta_Q\alpha[Z]&=\sqrt hNc_1\left[Z_{ij}\epsilon^{(i|kl}\mathrm{D}_kQ_l{}^{|j)}
+Q_{ij}\epsilon^{(i|kl}\mathrm{D}_kZ_l{}^{|j)}\right],\\
\delta_A\beta^{ij}[W]&=\sqrt h\epsilon^{(i|kl}
\left[2Nc_1W \mathrm{D}_kQ_l{}^{|j)}+\mathrm{D}_k(Nc_1W)Q_l{}^{|j)}\right],\\
\delta_Q\beta^{ij}[Z]&=2N\sqrt h(Z^{ij}-h^{ij}h^{kl}Z_{kl})\notag\\
&\quad+\sqrt h\epsilon^{(i|kl}
\left[2Nc_1A \mathrm{D}_kZ_l{}^{|j)}+\mathrm{D}_k(Nc_1A)Z_l{}^{|j)}\right].
\label{Fmultiplieroperator}
\end{align}
There is no term proportional to $\gamma$ in the $\alpha$ equation because $\delta_A\alpha[W]=0$. Setting the known source terms to zero and denoting the remaining scalar and tensor multiplier variations by $W$ and $Z_{ij}$ gives $\delta_Q\alpha[Z]=0$ and $\delta_A\beta[W]+\delta_Q\beta[Z]=0$. 
On the regular branch where the operator in Eq.~\eqref{Fmultiplieroperator} and its adjoint are invertible on the chosen boundary domain, Eq.~\eqref{Fsecondarypreserve} fixes $\gamma$ and $\nu^{(2)}_{ij}$ and produces no tertiary constraint. Appendix~\ref{app:regularity} supplies a constrained example and establishes the second-class rank on this branch.

The secondary constraints $\mathcal C_i\approx0$ generate spatial diffeomorphisms:
\begin{equation}
[\varPhi^I,\mathcal C[\zeta]]=\mathscr L_{\vec\zeta}\varPhi^I,\quad
[\varPi_I,\mathcal C[\zeta]]=\mathscr L_{\vec\zeta}\varPi_I,\quad
[\mathcal C[\zeta],\mathcal C[\eta]]=\mathcal C[[\zeta,\eta]].
\label{spatialclosure}
\end{equation}
Here $[\zeta,\eta]^i=\zeta^j\mathrm{D}_j\eta^i-\eta^j\mathrm{D}_j\zeta^i$ is the Lie bracket of the two spatial vectors. Hence preservation of $\mathcal C_i$ gives no further independent constraint, and $p_i,\mathcal C_i$ form the six first-class constraints.

We now summarize the  classification of all constraints in the theory:
\begin{itemize}
\item First class: $p_i\approx0$ and $\mathcal C_i\approx0$, giving six constraints.
\item Second class: $\phi_N\approx0,\ \varPi\approx0,\ \phi_h^{ij}\approx0,\ \varPi^{(1)}_{ij}\approx0,\ \pi\approx0,\ p_{(2)}^{ij}\approx0,\ \alpha\approx0,\ \beta^{ij}\approx0$, giving 28 constraints.
\end{itemize}
The number of DOFs is evaluated to be
\begin{equation}
\#_{\mathrm{dof}}^{(F)}=\frac{1}{2}(48-2\times 6-28)=4.
\label{Fcount}
\end{equation}
Therefore, there is an extra DOF in addition to the three DOFs in an ordinary SCG theory (without higher order time derivatives).

\section{The extrinsic-curvature-velocity interaction}
\label{sec:L}

\subsection{Primary constraints and total Hamiltonian}

We now focus on the monomial involving the Lie derivative of the extrinsic curvature (i.e., $\epsilon^{ijk}a_iK_j{}^l L_{kl}$) by setting $c_1=0$ and retaining $c_2$ in Eq.~\eqref{masteraction}. 
Again, we introduce an auxiliary field $Q_{ij}$ and impose $Q_{ij}=K_{ij}$ with the multiplier  $\lambda_{(1)}^{ij}$. 
The equivalent first-order Lagrangian is thus
\begin{align}
\widetilde L_L=\int \mathrm{d}^3x\big[&N\sqrt h({}^3R+Q_{ij}Q^{ij}-Q^2+c_3)
+B^{ij}(\dot Q_{ij}-\mathscr L_{\vec N}Q_{ij})\notag\\
&+\tfrac12\lambda_{(1)}^{ij}(\dot h_{ij}-\mathscr L_{\vec N}h_{ij})-N\lambda_{(1)}^{ij}Q_{ij}\big],
\label{Lauxaction}
\end{align}
where $B^{ij}$ is the coefficient of $\dot Q_{ij}$ and hence the momentum conjugate to $Q_{ij}$,
\begin{equation}
	B^{ij}\equiv \sqrt h c_2\epsilon^{rm(i}a_rQ_m{}^{j)}.
	\label{Bdef}
\end{equation}
The 22 configuration variables are $\varPhi^I=\{N,\allowbreak N^i,\allowbreak h_{ij},\allowbreak Q_{ij},\allowbreak \lambda_{(1)}^{ij}\}$. We denote their conjugate momenta as $\varPi_I=\{p,\allowbreak p_i,\allowbreak p_{(1)}^{ij},\allowbreak p_{(2)}^{ij},\allowbreak \varPi^{(1)}_{ij}\}$.

The definition $\varPi_I=\delta\widetilde L_L/\delta\dot\varPhi^I$ gives
\begin{equation}
p=0,\quad p_i=0,\quad p_{(1)}^{ij}=\tfrac12\lambda_{(1)}^{ij},\quad
p_{(2)}^{ij}=B^{ij},\quad\varPi^{(1)}_{ij}=0.
\label{Lmomenta}
\end{equation}
Accordingly, we have 22 primary constraints:
\begin{equation}
\begin{gathered}
p\approx0,\qquad p_i\approx0,\qquad
\phi_h^{ij}\equiv p_{(1)}^{ij}-\tfrac12\lambda_{(1)}^{ij}\approx0,\\
\phi_Q^{ij}\equiv p_{(2)}^{ij}-B^{ij}\approx0,\qquad\varPi^{(1)}_{ij}\approx0.
\end{gathered}
\label{Lprimary}
\end{equation}

The total Hamiltonian is
\begin{equation}
H_{\mathrm{T}}=H_L+\mathcal C[\vec N]+\int\mathrm{d}^3x\left(\xi p+\xi^ip_i+
\nu^{(1)}_{ij}\phi_h^{ij}+\nu^{(2)}_{ij}\phi_Q^{ij}+\mu^{ij}\varPi^{(1)}_{ij}\right),
\label{LHT}
\end{equation}
where the part independent of the shift vector is
\begin{equation}
H_L=\int\mathrm{d}^3x\,N\left[\lambda_{(1)}^{ij}Q_{ij}-\sqrt h({}^3R+Q_{ij}Q^{ij}-Q^2+c_3)\right].
\label{H0}
\end{equation}
The subscript $L$ identifies the interaction containing $L_{ij}$, in parallel with $H_F$ in Sec.~\ref{sec:F}. The generator $\mathcal C[\zeta]$ of spatial diffeomorphisms is given by Eqs.~\eqref{spatialgeneratorsmeared} and \eqref{spatialgenerator}, with the canonical pairs $(A,\pi)$ and $(\lambda,\varPi)$ omitted.

\subsection{Primary preservation and secondary constraints}

Similar to the discussion in Sec.~\ref{sec:F}, in order to determine the preservation of the primary constraints (Eq.~\eqref{preservationrule} applies with $H_F$ replaced by $H_L$), it is necessary to evaluate the Poisson brackets among all primary constraints.
The complete primary Poisson-bracket matrix $\mathbb P_L(\vec{x},\vec{y})$ is given by
\begin{equation}
	\begingroup\setlength{\arraycolsep}{3pt}
	\begin{array}{c|ccccc}
		[\cdot,\cdot]&p(\vec{y})&p_j(\vec{y})&\phi_h^{kl}(\vec{y})&\phi_Q^{kl}(\vec{y})&\varPi^{(1)}_{kl}(\vec{y})\\\hline
		p(\vec{x})&0&0&0&\dfrac{\delta B^{kl}(\vec{y})}{\delta N(\vec{x})}&0\\
		p_i(\vec{x})&0&0&0&0&0\\
		\phi_h^{ij}(\vec{x})&0&0&0&\dfrac{\delta B^{kl}(\vec{y})}{\delta h_{ij}(\vec{x})}&-\tfrac12\delta^{ij}_{kl}\delta^3(\vec{x}-\vec{y})\\
		\phi_Q^{ij}(\vec{x})&-\dfrac{\delta B^{ij}(\vec{x})}{\delta N(\vec{y})}&0&-\dfrac{\delta B^{ij}(\vec{x})}{\delta h_{kl}(\vec{y})}&-2\sqrt h c_2\epsilon^{rm(i}a_r\delta_m^{(k}h^{l)j)}\delta^3(\vec{x}-\vec{y})&0\\
		\varPi^{(1)}_{ij}(\vec{x})&0&0&\tfrac12\delta_{ij}^{kl}\delta^3(\vec{x}-\vec{y})&0&0
	\end{array}.
	\endgroup
	\label{Lprimarymatrix}
\end{equation}
The $N$ and $h$ derivatives of $B^{ij}$ enter the mixed brackets in Eq.~\eqref{Lprimarymatrix}, while its $Q$ derivative enters the $\phi_Q$--$\phi_Q$ bracket through antisymmetrization. In the notation of Eq.~\eqref{variationdefinition}, with $W=\delta N$, $Y_{ij}=\delta h_{ij}$, and $Z_{ij}=\delta Q_{ij}$, these variations are
\begin{align}
	\delta_N B^{ij}[W]&=\sqrt h\epsilon^{rm(i}\mathrm{D}_r(c_2W/N)Q_m{}^{j)},\label{BN}\\
	\delta_h B^{ij}[Y]&=-\sqrt h c_2\epsilon^{rm(i}a_rY^{j)k}Q_{mk},\label{Bh}\\
	\delta_Q B^{ij}[Z]&=\sqrt h c_2\epsilon^{rm(i}a_rZ_m{}^{j)}.\label{BQ}
\end{align}
Finally, antisymmetrizing the two $Q$ derivatives gives the factor $-2$ in the last tensor entry of Eq.~\eqref{Lprimarymatrix}. Explicitly,
\begin{align}
	[p(\vec{x}),\phi_Q[Z]]&=-\sqrt h\frac{c_2}{N}\epsilon^r{}_{mn}\mathrm{D}_r(Q^{mi}Z_i{}^n),\notag\\
	[\phi_Q^{ij}(\vec{x}),p[W]]&=-\sqrt h\epsilon^{rm(i}\mathrm{D}_r(c_2W/N)Q_m{}^{j)},\notag\\
	[\phi_Q^{ij}(\vec{x}),\phi_Q[Z]]&=-2\sqrt h c_2\epsilon^{rm(i}a_rZ_m{}^{j)}.
	\label{Lprimaryactions}
\end{align}
The first equation follows by integrating Eq.~\eqref{BN} once, and the second is its negative transpose. 
Equations~\eqref{BN}--\eqref{Lprimaryactions} thus specify every nonvanishing entry in the primary Poisson-bracket matrix.

Preservation of $\varPi^{(1)}_{ij}$ fixes $\nu^{(1)}_{ij}$, while preservation of $\phi_h^{ij}$ expresses $\mu^{ij}$ in terms of the remaining $\nu^{(2)}_{ij}$:
\begin{align}
\nu^{(1)}_{ij}&=2NQ_{ij},\\
\mu^{ij}&=-2\frac{\delta H_L}{\delta h_{ij}}
-2\sqrt h c_2\nu^{(2)}_{kl}\epsilon^{rmk}a_rh^{l(i}Q_m{}^{j)}.
\label{Lfixedaux}
\end{align}
Preservation of the three $p_i$ gives $\mathcal C_i\approx0$. Preservation of $p\approx0$ and the six $\phi_Q^{ij}\approx0$, respectively, gives the remaining scalar and tensor equations. Substituting Eq.~\eqref{Lfixedaux} and $\lambda_{(1)}^{ij}\approx2p_{(1)}^{ij}$, they read
\begin{align}
0={}&-2p_{(1)}^{ij}Q_{ij}+\sqrt h[{}^3R+Q_{ij}Q^{ij}-Q^2+c_3+Nc_{3N}]
-\sqrt h\frac{c_2}{N}\epsilon^r{}_{mn}\mathrm{D}_r(Q^{mi}\nu^{(2)}_i{}^n),\label{LpPreserve}\\
0={}&-\sqrt h \frac{\partial c_2}{\partial t}\epsilon^{rm(i}a_rQ_m{}^{j)}
-2Np_{(1)}^{ij}+2N\sqrt h(Q^{ij}-Qh^{ij})\notag\\
&+2N\sqrt h c_2\epsilon^{rm(i}a_rQ^{j)k}Q_{mk}
-\sqrt h\epsilon^{rm(i}\mathrm{D}_r(c_2\xi/N)Q_m{}^{j)}
-2\sqrt h c_2\epsilon^{rm(i}a_r\nu^{(2)}_m{}^{j)}.
\label{LQPreserve}
\end{align}

Contracting Eq.~\eqref{LQPreserve} with $h_{ij}$ eliminates every multiplier term and gives an additional secondary constraint. To identify the corresponding primary direction, define
\begin{equation}
T_0\equiv h_{ij}\phi_Q^{ij}=h_{ij}p_{(2)}^{ij}\approx0.
\label{traceprimary}
\end{equation}
Here $h_{ij}B^{ij}=Q_{ij}B^{ij}=0$. In the primary matrix \eqref{Lprimarymatrix}, the trace test tensor $Z_{ij}=zh_{ij}$ annihilates the $p$ and $\phi_Q$ rows. Its contribution to the $\phi_h$ row is canceled by the multiplier component $-2zB^{ij}$ along $\varPi^{(1)}_{ij}$. Thus the primary combination
\begin{equation}
\widetilde T_0\equiv T_0-2B^{ij}\varPi^{(1)}_{ij}
\label{tracenullprimary}
\end{equation}
is a null direction on the primary constraint surface, without any additional restriction on the fields. After imposing the auxiliary primary constraints, it reduces to $T_0$. Taking the trace of Eq.~\eqref{LQPreserve}, or preserving $T_0$ after Eq.~\eqref{Lfixedaux}, gives
\begin{equation}
\dot T_0\approx-2N\left(h_{ij}p_{(1)}^{ij}+2\sqrt h Q\right)\approx0.
\label{tracepreservation}
\end{equation}
Since $N>0$, this generates
\begin{equation}
S\equiv h_{ij}p_{(1)}^{ij}+2\sqrt h Q\approx0,
\label{Ssecondary}
\end{equation}
in addition to the three secondary constraints $\mathcal C_i\approx0$.

\subsection{Secondary preservation and number of degrees of freedom}

We now turn to the preservation conditions for the secondary constraints $S\approx 0$ and $\mathcal{C}_{i}$.

The Poisson brackets involving the secondary constraint $S\approx 0$ are
\begin{align}
[T_0(\vec{x}),S(\vec{y})]&=\left[T_0(\vec{x})-6\sqrt{h(\vec{x})}\right]\delta^3(\vec{x}-\vec{y})
\approx-6\sqrt{h(\vec{x})}\delta^3(\vec{x}-\vec{y}),\label{TSbracket}\\
[S(\vec{x}),\phi_h^{kl}(\vec{y})]&=\left[p_{(1)}^{kl}+\sqrt h Qh^{kl}-2\sqrt h Q^{kl}\right](\vec{x})\delta^3(\vec{x}-\vec{y}),\notag\\
[S(\vec{x}),\phi_Q^{kl}(\vec{y})]&=\left[2\sqrt h h^{kl}-B^{kl}\right](\vec{x})\delta^3(\vec{x}-\vec{y}),\notag\\
[S(\vec{x}),p(\vec{y})]&=[S(\vec{x}),p_i(\vec{y})]=0,\notag\\
[S(\vec{x}),\varPi^{(1)}_{kl}(\vec{y})]&=[S(\vec{x}),S(\vec{y})]=0.
\label{Sbrackets}
\end{align}
In particular,
\begin{align}
[S(\vec{x}),p_{(2)}^{kl}(\vec{y})]&=2\sqrt{h(\vec{x})}h^{kl}(\vec{x})\delta^3(\vec{x}-\vec{y}),\notag\\
[S(\vec{x}),B^{kl}(\vec{y})]&=B^{kl}(\vec{x})\delta^3(\vec{x}-\vec{y}).
\label{Sbracketderivation}
\end{align}

Using $[S(\vec{x}),H_L]=-h_{ij}(\vec{x})\delta H_L/\delta h_{ij}(\vec{x})$ and substituting $\nu^{(1)}_{ij}=2NQ_{ij}$, the single scalar preservation equation for $S$ is
\begin{equation}
0=V_0+(2\sqrt h h^{ij}-B^{ij})\nu^{(2)}_{ij},\label{S_preserve_L}
\end{equation}
where $V_0$ is given by
\begin{equation}
V_0=2Np_{(1)}^{ij}Q_{ij}
+N\sqrt h\left(\frac{1}{2}{}^3R-\frac{9}{2}Q_{ij}Q^{ij}+\frac{5}{2}Q^2+\frac{3}{2}c_3\right)
-2\sqrt h \mathrm{D}^2N.
\label{V0}
\end{equation}
We decompose $\nu^{(2)}_{ij}$ into its trace and trace-free parts by $\nu^{(2)}_{ij}=z_0h_{ij}+\nu^{(2),\mathrm{TF}}_{ij}$ such that $h^{ij}\nu^{(2),\mathrm{TF}}_{ij}=0$. The coefficient of the trace part is then
\begin{equation}
z_0=-\frac{V_0-B^{ij}\nu^{(2),\mathrm{TF}}_{ij}}{6\sqrt h}.
\label{tracefixL}
\end{equation}
Again, the three secondary constraints $\mathcal C_i\approx0$ are automatically preserved due to Eq.~\eqref{spatialclosure}. The six equations for $\xi$ and $\nu^{(2),\mathrm{TF}}_{ij}$ are the scalar primary equation \eqref{LpPreserve} and the five trace-free components of the tensor primary equation \eqref{LQPreserve}. On a regular boundary domain, these fix the six multipliers, and Eq.~\eqref{tracefixL} fixes the remaining trace. The required rank is justified in Appendix~\ref{app:regularity}. Thus no tertiary constraint arises on this branch.

Let $\boldsymbol{\varphi}_L=(p,p_i,\phi_h^{ij},\phi_Q^{ij},\varPi^{(1)}_{ij})$ denote the primary constraints in Eq.~\eqref{Lprimary}. The full Poisson-bracket matrix among all the primary and secondary constraints is
\begin{equation}
	\begin{array}{c|ccc}
		[\cdot,\cdot]&\boldsymbol{\varphi}_L(\vec{y})&S(\vec{y})&\mathcal C_j(\vec{y})\\\hline
		\boldsymbol{\varphi}_L(\vec{x})&\mathbb P_L&\mathbb B_L&0\\
		S(\vec{x})&-\mathbb B_L^{\mathrm T}&0&0\\
		\mathcal C_i(\vec{x})&0&0&0
	\end{array}, \label{Lfullmatrix}
\end{equation}
where the block of the primary constraints is given by Eq.~\eqref{Lprimarymatrix}, and
\begin{equation}
\mathbb B_L(\vec{x},\vec{y})=
\begin{array}{c|c}
[\cdot,\cdot]&S(\vec{y})\\\hline
p(\vec{x})&0\\
p_i(\vec{x})&0\\
\phi_h^{ij}(\vec{x})&-\dfrac{\delta S(\vec{y})}{\delta h_{ij}(\vec{x})}\\
\phi_Q^{ij}(\vec{x})&-\bigl(2\sqrt{h}h^{ij}-B^{ij}\bigr)(\vec{x})\delta^3(\vec{x}-\vec{y})\\
\varPi^{(1)}_{ij}(\vec{x})&0
\end{array}.
\end{equation}
The derivative $\delta_h S$ and the other nonzero entries of $\mathbb B_L$ are given in Eq.~\eqref{Sbrackets}. 
In \eqref{Lfullmatrix}, only the blocks involving $\mathcal C_i$ are weak zeros. The nonzero bracket \eqref{TSbracket} pairs $S$ with the previously null primary direction $\widetilde T_0$ in Eq.~\eqref{tracenullprimary}, or equivalently with $T_0$ after eliminating the auxiliary constraints.

We are now ready to classify the constraints.

The six pairs $(\phi_h^{ij},\varPi^{(1)}_{ij})$ contribute rank 12. After eliminating them, the seven nonspatial primaries $(p,\phi_Q^{ij})$ have rank six on the regular branch specified above, with the trace as their only null direction. Since $[T_0,S]\approx-6\sqrt h\,\delta^3$, adding $S$ raises the rank by two. Hence the primary rank is $12+6=18$ and the full rank is $12+8=20$. The full set of 26 constraints is therefore classified as
\begin{itemize}
\item First class: $p_i\approx0$ and $\mathcal C_i\approx0$, giving six constraints.
\item Second class: $p\approx0,\ \phi_h^{ij}\approx0,\ \varPi^{(1)}_{ij}\approx0,\ \phi_Q^{ij}\approx0,\ S\approx0$, giving 20 constraints.
\end{itemize}
Thus the total number of degrees of freedom is given by
\begin{equation}
\#_{\mathrm{dof}}^{(L)}=\frac{1}{2}(44-2\times6-20)=6.
\label{Lcount}
\end{equation}
Thus, the existence of the monomial involving the Lie derivative of the extrinsic curvature introduces 3 extra DOFs in addition to the 3 DOFs in an ordinary SCG theory.

\section{Both higher-time-derivative interactions}
\label{sec:both}

We now retain both higher-time-derivative interactions in Eq.~\eqref{masteraction} and determine whether their combined constraint structure reduces the number of degrees of freedom found in Sec.~\ref{sec:L}.

\subsection{Primary constraints and total Hamiltonian}

For the full action \eqref{masteraction}, the equivalent first-order Lagrangian is
\begin{align}
\widetilde L_{F+L}=\int\mathrm d^3x\big[&N\sqrt h({}^3R+Q_{ij}Q^{ij}-Q^2+c_3)
+C(\dot N-N^i\mathrm D_iN)+B^{ij}(\dot Q_{ij}-\mathscr L_{\vec N}Q_{ij})\notag\\
&+\tfrac12\lambda_{(1)}^{ij}(\dot h_{ij}-\mathscr L_{\vec N}h_{ij})-N\lambda_{(1)}^{ij}Q_{ij}\big],
\label{bothauxaction}
\end{align}
where
\begin{equation}
C\equiv\sqrt h\frac{c_1}{N}\chi,
\label{Cdef}
\end{equation}
and $B^{ij}$ is unchanged from Eq.~\eqref{Bdef}. The lapse-velocity term contains no $\dot Q_{ij}$ and therefore does not alter this coefficient.
The 22 configuration variables are again $\varPhi^I=\{N,\allowbreak N^i,\allowbreak h_{ij},\allowbreak Q_{ij},\allowbreak \lambda_{(1)}^{ij}\}$, with conjugate momenta $\varPi_I=\{p,\allowbreak p_i,\allowbreak p_{(1)}^{ij},\allowbreak p_{(2)}^{ij},\allowbreak \varPi^{(1)}_{ij}\}$. Writing the first-order Lagrangian as $\widetilde L_{F+L}$, the definition $\varPi_I=\delta\widetilde L_{F+L}/\delta\dot\varPhi^I$ gives
\begin{equation}
p=C,\quad p_i=0,\quad p_{(1)}^{ij}=\lambda_{(1)}^{ij}/2,\quad
p_{(2)}^{ij}=B^{ij},\quad\varPi^{(1)}_{ij}=0,
\label{bothmomenta}
\end{equation}
where $C$ and $B^{ij}$ are given in Eqs.~\eqref{Cdef} and \eqref{Bdef}, respectively.
Thus we have 22 primary constraints
\begin{equation}
\begin{gathered}
\phi_N\equiv p-C\approx0,\qquad p_i\approx0,\qquad
\phi_h^{ij}\equiv p_{(1)}^{ij}-\tfrac12\lambda_{(1)}^{ij}\approx0,\\
\phi_Q^{ij}\equiv p_{(2)}^{ij}-B^{ij}\approx0,\qquad\varPi^{(1)}_{ij}\approx0.
\end{gathered}
\label{bothprimary}
\end{equation}
Both parity-odd velocity terms cancel in the Legendre transform. The total Hamiltonian is
\begin{equation}
H_{\mathrm T}=H_{F+L}+\mathcal C[\vec N]+\int\mathrm d^3x\left(
\xi\phi_N+\xi^ip_i+\nu^{(1)}_{ij}\phi_h^{ij}+\nu^{(2)}_{ij}\phi_Q^{ij}
+\mu^{ij}\varPi^{(1)}_{ij}\right),
\label{bothHT}
\end{equation}
with
\begin{equation}
H_{F+L}=\int\mathrm d^3x\,N\left[\lambda_{(1)}^{ij}Q_{ij}
-\sqrt h({}^3R+Q_{ij}Q^{ij}-Q^2+c_3)\right]=H_L.
\label{bothH}
\end{equation}

The Poisson-bracket matrix of all primary constraints $\mathbb P_{F+L}(\vec{x},\vec{y})$ is
\begin{equation}
\begingroup\setlength{\arraycolsep}{3pt}
\begin{array}{c|ccccc}
[\cdot,\cdot]&\phi_N(\vec{y})&p_j(\vec{y})&\phi_h^{kl}(\vec{y})&\phi_Q^{kl}(\vec{y})&\varPi^{(1)}_{kl}(\vec{y})\\\hline
\phi_N(\vec{x})&0&0&-\dfrac{\delta C(\vec{x})}{\delta h_{kl}(\vec{y})}&M^{kl}(\vec{x},\vec{y})&0\\
p_i(\vec{x})&0&0&0&0&0\\
\phi_h^{ij}(\vec{x})&\dfrac{\delta C(\vec{y})}{\delta h_{ij}(\vec{x})}&0&0&\dfrac{\delta B^{kl}(\vec{y})}{\delta h_{ij}(\vec{x})}&-\tfrac12\delta^{ij}_{kl}\delta^3(\vec{x}-\vec{y})\\
\phi_Q^{ij}(\vec{x})&-M^{ij}(\vec{y},\vec{x})&0&-\dfrac{\delta B^{ij}(\vec{x})}{\delta h_{kl}(\vec{y})}&-2\sqrt h c_2\epsilon^{rm(i}a_r\delta_m^{(k}h^{l)j)}\delta^3(\vec{x}-\vec{y})&0\\
\varPi^{(1)}_{ij}(\vec{x})&0&0&\tfrac12\delta_{ij}^{kl}\delta^3(\vec{x}-\vec{y})&0&0
\end{array},
\endgroup
\label{bothprimarymatrix}
\end{equation}
where
\begin{equation}
M^{ij}(\vec{x},\vec{y})=\frac{\delta B^{ij}(\vec{y})}{\delta N(\vec{x})}
-\frac{\delta C(\vec{x})}{\delta Q_{ij}(\vec{y})}.
\label{mixedkerneldefinition}
\end{equation} The bracket $[\phi_N(\vec{x}),\phi_N(\vec{y})]$ vanishes because $C$ has no spatial derivative of $N$. The functional derivatives involving $B^{ij}$ are the same as those in Eqs.~\eqref{BN}--\eqref{BQ}. The functional derivative  $\delta_hC[Y]=(c_1/N)\delta_h(\sqrt h\chi)[Y]$ is evaluated in Eq.~\eqref{metriclocal}.

To calculate $[\phi_N(\vec{x}),\phi_Q[Z]]$, combine the first line of Eq.~\eqref{Lprimaryactions} with $-\delta_QC[Z]$ from Eq.~\eqref{localQvariation}. The identities
\begin{align}
\epsilon^r{}_{mn}(\mathrm{D}_rQ^{mi})Z_i{}^n&=Z_{ij}\epsilon^{(i|kl}\mathrm{D}_kQ_l{}^{|j)},\notag\\
\epsilon^r{}_{mn}Q^{mi}\mathrm{D}_rZ_i{}^n&=-Q_{ij}\epsilon^{(i|kl}\mathrm{D}_kZ_l{}^{|j)}
\end{align}
then give
\begin{align}
[\phi_N(\vec{x}),\phi_Q[Z]]=-\frac{\sqrt h}{N}\big[&(c_1+c_2)Z_{ij}\epsilon^{(i|kl}\mathrm{D}_kQ_l{}^{|j)}
+(c_1-c_2)Q_{ij}\epsilon^{(i|kl}\mathrm{D}_kZ_l{}^{|j)}\big],\label{mixedaction}\\
[\phi_Q^{ij}(\vec{x}),\phi_N[W]]=\sqrt h\epsilon^{(i|kl}\bigg[&2\frac{c_1}{N}W \mathrm{D}_kQ_l{}^{|j)}
+\mathrm{D}_k\left(\frac{c_1-c_2}{N}W\right)Q_l{}^{|j)}\bigg].
\label{mixedtranspose}
\end{align}
Note that all the fields on the right-hand-sides are evaluated at $\vec{x}$. 

\subsection{Primary preservation and secondary constraints}

We now examine the preservation of primary constraints.
The general preservation equation is the same as Eq.~\eqref{preservationrule}, by simply changing $H_{F}$ to  $H_L$ with the primary constraints \eqref{bothprimary}. 

The auxiliary primary constraints obey
\begin{align}
0&\approx\dot\varPi^{(1)}_{ij}=-NQ_{ij}+\tfrac12\nu^{(1)}_{ij},\notag\\
0&\approx\dot\phi_h^{ij}=-\frac{\delta H_L}{\delta h_{ij}}-\tfrac12\mu^{ij}
+\frac{\delta}{\delta h_{ij}}\int\mathrm d^3y\,
\left(\xi C+\nu^{(2)}_{kl}B^{kl}\right).
\label{bothauxpreserve}
\end{align}
Thus $\nu^{(1)}_{ij}=2NQ_{ij}$, while $\mu^{ij}$ is determined in terms of the remaining multipliers $\xi$ and $\nu^{(2)}_{ij}$:
\begin{align}
\mu^{ij}={}&-2\frac{\delta H_L}{\delta h_{ij}}
+2\sqrt h\left[-\frac{c_1\xi}{N}\epsilon^{abc}Q_a{}^{(i}\mathrm{D}_bQ_c{}^{j)}
+\mathrm{D}_l\left(\frac{c_1\xi}{N}\epsilon^{a(i|c}Q_a{}^lQ_c{}^{|j)}\right)\right]\notag\\
&-2\sqrt h c_2\nu^{(2)}_{kl}\epsilon^{rmk}a_rh^{l(i}Q_m{}^{j)},
\label{bothmu}
\end{align}
These twelve equations generate no secondary constraint.
Again, the preservation of $p_i$ yields the secondary constraints $\mathcal C_i\approx0$. The remaining seven preservation equations for the primary constraints are
\begin{align}
	0={}&-2p_{(1)}^{ij}Q_{ij}+\sqrt h[{}^3R+Q_{ij}Q^{ij}-Q^2+c_3+Nc_{3N}]
	-\sqrt h\frac{c_{1t}}{N}\chi+2\sqrt h c_1Q\chi\notag\\
	&-\frac{\sqrt h}{N}\big[(c_1+c_2)\nu^{(2)}_{ij}\epsilon^{(i|kl}\mathrm{D}_kQ_l{}^{|j)}
	+(c_1-c_2)Q_{ij}\epsilon^{(i|kl}\mathrm{D}_k\nu^{(2)}_l{}^{|j)}\big],\label{bothscalar}
\end{align}
and
\begin{align}
0={}&-\sqrt h \frac{\partial c_2}{\partial t}\epsilon^{rm(i}a_rQ_m{}^{j)}-2Np_{(1)}^{ij}+2N\sqrt h(Q^{ij}-Qh^{ij})
+2N\sqrt h c_2\epsilon^{rm(i}a_rQ^{j)k}Q_{mk}\notag\\
&+\sqrt h\epsilon^{(i|kl}\left[2\frac{c_1\xi}{N}\mathrm{D}_kQ_l{}^{|j)}
+\mathrm{D}_k\left(\frac{c_1-c_2}{N}\xi\right)Q_l{}^{|j)}\right]
-2\sqrt h c_2\epsilon^{rm(i}a_r\nu^{(2)}_m{}^{j)}.
\label{bothtensor}
\end{align}
The simplification of the metric source in the first equation uses
\begin{equation}
\delta_h(\sqrt h\chi)[2NQ]=-2N\sqrt h Q\chi.
\label{metricvelocity}
\end{equation}
We refer to Appendix~\ref{app:variations} for more details on evaluating Poisson brackets.

As in Sec.~\ref{sec:L}, the primary combination $\widetilde T_0$ in Eq.~\eqref{tracenullprimary}, associated with the trace $h_{ij}\phi_Q^{ij}$, remains a null direction of the primary Poisson-bracket matrix. Indeed, both terms in Eq.~\eqref{mixedaction} vanish for a trace test tensor $Z_{ij}=\zeta h_{ij}$, where $\zeta$ is an arbitrary test function. Taking the trace of Eq.~\eqref{bothtensor} therefore gives the fourth secondary constraint
\begin{equation}
S=h_{ij}p_{(1)}^{ij}+2\sqrt h Q\approx0,
\label{bothsecondary}
\end{equation}
in addition to $\mathcal C_i\approx0$. The scalar equation \eqref{bothscalar} and the five trace-free components of Eq.~\eqref{bothtensor} determine six multipliers on the regular branch, leaving the trace of $\nu^{(2)}_{ij}$ to be fixed by preservation of $S$.

\subsection{Secondary preservation and number of degrees of freedom}

The only additional bracket involving $S$ is
\begin{equation}
[S(\vec{x}),\phi_N(\vec{y})]=-C(\vec{x})\delta^3(\vec{x}-\vec{y}),
\label{SC}
\end{equation}
which follows from $\delta_h(\sqrt h\chi)[\zeta h]=-\zeta\sqrt h\chi$. The other brackets involving $S$ remain those in Eqs.~\eqref{Sbrackets} and \eqref{TSbracket}. The preservation of $S\approx0$ gives the single scalar equation
\begin{equation}
0=V_0-C\xi+(2\sqrt h h^{ij}-B^{ij})\nu^{(2)}_{ij},
\label{bothS}
\end{equation}
where $V_0$ is defined in Eq.~\eqref{V0}. Writing $\nu^{(2)}_{ij}=z_0h_{ij}+\nu^{(2),\mathrm{TF}}_{ij}$, this equation fixes the trace coefficient:
\begin{equation}
z_0=-\frac{V_0-C\xi-B^{ij}\nu^{(2),\mathrm{TF}}_{ij}}{6\sqrt h}.
\label{bothtracefix}
\end{equation}
Here $z_0$ is a multiplier component, whereas $\zeta$ above is only a test function. The three $\mathcal C_i$ are preserved by Eq.~\eqref{spatialclosure}.

We denote $\boldsymbol{\varphi}_{F+L}=(\phi_N,p_i,\phi_h^{ij},\phi_Q^{ij},\varPi^{(1)}_{ij})$ as the primary constraints in Eq.~\eqref{bothprimary}. The complete Poisson-bracket matrix of all constraints is
\begin{equation}
	\begin{array}{c|ccc}
		[\cdot,\cdot]&\boldsymbol{\varphi}_{F+L}(\vec{y})&S(\vec{y})&\mathcal C_j(\vec{y})\\\hline
		\boldsymbol{\varphi}_{F+L}(\vec{x})&\mathbb P_{F+L}&\mathbb B_{F+L}&0\\
		S(\vec{x})&-\mathbb B_{F+L}^{\mathrm T}&0&0\\
		\mathcal C_i(\vec{x})&0&0&0
	\end{array}, \label{bothfullmatrix}
\end{equation}
where
\begin{equation}
\mathbb B_{F+L}(\vec{x},\vec{y})=
\begin{array}{c|c}
[\cdot,\cdot]&S(\vec{y})\\\hline
\phi_N(\vec{x})&C(\vec{x})\delta^3(\vec{x}-\vec{y})\\
p_i(\vec{x})&0\\
\phi_h^{ij}(\vec{x})&-\dfrac{\delta S(\vec{y})}{\delta h_{ij}(\vec{x})}\\
\phi_Q^{ij}(\vec{x})&-\bigl(2\sqrt{h}h^{ij}-B^{ij}\bigr)(\vec{x})\delta^3(\vec{x}-\vec{y})\\
\varPi^{(1)}_{ij}(\vec{x})&0
\end{array}.
\end{equation}
The block of the primary constraints is given by Eq.~\eqref{bothprimarymatrix}. Relative to Eq.~\eqref{Lfullmatrix}, the additional secondary bracket is $[\phi_N(\vec{x}),S(\vec{y})]=C(\vec{x})\delta^3(\vec{x}-\vec{y})$, in agreement with Eq.~\eqref{SC}. 
The zero blocks involving $\mathcal C_i$ in Eq.~\eqref{bothfullmatrix} are weak equalities. All other displayed entries are exact.

The brackets $[\phi_N,\phi_Q^{ij}]$ must also exclude an additional null direction independent of $\widetilde T_0$. Appendix~\ref{app:regularity} establishes a regular domain for the scalar equation \eqref{bothscalar} and the five trace-free tensor equations from Eq.~\eqref{bothtensor}. On this branch they fix $\xi$ and the five components of $\nu^{(2),\mathrm{TF}}_{ij}$, while Eq.~\eqref{bothtracefix} fixes $z_0$. The primary rank is $12+6=18$, and the nonzero bracket \eqref{TSbracket} raises the full rank to 20. No tertiary constraint arises. The 26 constraints are therefore classified as
\begin{itemize}
\item First class: $p_i\approx0$ and $\mathcal C_i\approx0$, giving six constraints.
\item Second class: $\phi_N\approx0,\ \phi_h^{ij}\approx0,\ \varPi^{(1)}_{ij}\approx0,\ \phi_Q^{ij}\approx0,\ S\approx0$, giving 20 constraints.
\end{itemize}
As a result, the total number of degrees of freedom is counted to be
\begin{equation}
\#_{\mathrm{dof}}^{(F+L)}=\frac{1}{2}(44-2\times6-20)=6.
\label{bothcount}
\end{equation}
As a result, the extrinsic-curvature-velocity interaction gives six degrees of freedom, with or without the lapse-velocity interaction. Neither equal nor opposite nonzero coefficients produce an additional degeneracy valid for general field configurations, as shown in Appendix~\ref{app:regularity}.

\section{Conclusion}
\label{sec:conclusion}

In this work, we investigated the complete parity-violating scalar-tensor basis up to $d=4$ and the degrees of freedom of its higher-time-derivative interactions by combining the SCG--GST correspondence with a Hamiltonian constraint analysis. 
The classification in Secs.~\ref{subsec:GSTbasis} and \ref{subsec:SCGbasis} separates pointwise linear independence from independence under integrations by parts. Including all derivative partitions gives one PV direction at $d=3$ and thirteen at $d=4$ in both the GST and SCG descriptions, Eqs.~\eqref{pointwisecount} and \eqref{SCGpointwisecount}. 

The six integration-by-parts reductions in Eqs.~\eqref{weightedIBP-1}--\eqref{weightedIBP-3} and \eqref{higherIBP1}--\eqref{higherIBP3} give the SCG action basis \eqref{PVactionbasis}, corresponding to the GST basis \eqref{GSTactionbasis}. Both contain one $d=3$ and seven $d=4$ representatives. The bidirectional coefficient maps \eqref{forwardcoeff} and \eqref{inversecoeff}, together with the explicit divergence in Eq.~\eqref{fullactiondictionary-1}, preserve coefficient derivatives and induced lower-grade terms. The subset \eqref{safeSCG} and its covariant representatives \eqref{safeL2}--\eqref{safeL7}, together with $\mathcal G_0$ in Eq.~\eqref{Gzero}, provide a complete basis of one $d=3$ and five $d=4$ combinations without higher time derivatives in unitary gauge.

We then perform a Hamiltonian constraint analysis to examine whether the two SCG monomials with higher-order time derivatives propagate extra degrees of freedom (DOFs).
For the action \eqref{masteraction}, with the standard Einstein-Hilbert kinetic term and a lapse-dependent potential, Sec.~\ref{sec:F} finds four DOFs for the lapse-velocity interaction, Eq.~\eqref{Fcount}, establishing that this term introduces one additional mode. Section~\ref{sec:L} finds six DOFs for the extrinsic-curvature-velocity interaction, Eq.~\eqref{Lcount}, corresponding to three additional modes. The primary combination \eqref{tracenullprimary} generates the secondary constraint \eqref{Ssecondary}, whose nonzero bracket \eqref{TSbracket} completes a second-class pair.

In Sec.~\ref{sec:both}, we extend the analysis in the presence of both higher-time-derivative terms. We also find that generally the theory propagates six DOFs as in Eq.~\eqref{bothcount}. 
In particular, there is no universal cancellation of the additional modes by equal or opposite nonzero coefficients within Eq.~\eqref{masteraction}. 

The results presented in this work specify how to construct the most general parity-violating scalar-tensor theory with three DOFs up to $d=4$, whose parity-violating Lagrangian is a linear combination of six GST polynomials with arbitrary coefficient functions, as identified in Sec.~\ref{sec:basis} and Eqs.~\eqref{Gzero} and \eqref{safeL2}--\eqref{safeL7}. The combination of a complete operator classification, coefficient-dependent integrations by parts, and Hamiltonian constraint analysis also provides a systematic method that can be extended to $d=5$.

%%
%%%
\begin{acknowledgments}
	This work is supported by the National Natural Science Foundation of China (NSFC) under Grant No. 12475068 and the Guangdong Basic and Applied Basic Research Foundation under Grant No. 2025A1515012977.
	We used ChatGPT (GPT-6) to cross-check integration-by-parts reductions and intermediate steps in the Hamiltonian constraint analysis. All results and conclusions were independently verified by the authors.
\end{acknowledgments}

\appendix

\section{Algebraic reduction and independence under integrations by parts}
\label{app:algebra}

\subsection{Algebraic identities}
The identities in Eqs.~\eqref{sevenidentities-1}--\eqref{sevenidentities-7} reduce the original seventeen candidates to ten. The following component decomposition proves completeness and independence of that ten-dimensional space.
The scalar derivative decomposition is
\begin{equation}
	\nabla_a\phi_{bc}=\nabla_{(a}\phi_{bc)}
	+\frac13\left(R_{cdab}+R_{bdac}\right)\phi^d.
	\label{thirdsplit}
\end{equation}
For the SCG reduction, we use
\begin{equation}
	\phi_{ab}=\sigma(-K_{ab}+u_aa_b+u_ba_a-Fu_au_b),
	\label{scalarseconddecomposition}
\end{equation}
and the Gauss-Codazzi relations
\begin{align}
	R_{0i0j}&=-L_{ij}+K_i{}^mK_{mj}+\mathrm{D}_ia_j+a_ia_j,\notag\\
	R_{0ijk}&=\mathrm{D}_kK_{ij}-\mathrm{D}_jK_{ik},\notag\\
	R_{ijkl}&={}^3R_{ijkl}+K_{ik}K_{jl}-K_{il}K_{jk},
	\label{GC}
\end{align}
Here $0$ denotes the unit normal in an adapted orthonormal frame.

Equations~\eqref{thirdsplit}--\eqref{GC}, the Bianchi identities, and the dictionary \eqref{rawdictionary} also verify the seven relations quoted in the main text.

An explicit component proof uses an oriented orthonormal frame at a single point with $\phi_a=(\sigma,0,0,0)$. All indices $i,j,k,l,m$ below range over the three spatial directions and are summed with $\delta_{ij}$. The time label $0$ refers only to this frame. The second scalar derivative, fully symmetric third scalar derivative, and algebraic curvature can be specified independently. The curvature components $R_{0i0j}$ and $\sum_mR_{mimj}$ are independently symmetric. The nine contractions $\tfrac12\epsilon^{jkl}R_{0ikl}$ have zero trace but are not necessarily symmetric, giving the remaining eight curvature components.

The ten possible PV contractions can be written without additional tensor definitions. Their expressions in terms of the GST basis are
\begin{align}
\frac{1}{2\sigma^2}\epsilon^{jkl}R_{0ikl}\phi_{im}\phi_{mj}
&=\tfrac12(\mathcal G_1+\mathcal G_4),\label{inversecomponents-1}
\\
\frac{1}{2\sigma^2}\left(\sum_m\phi_{mm}\right)\epsilon^{jkl}R_{0ikl}\phi_{ij}
&=\mathcal G_2+\tfrac12(\mathcal G_1+\mathcal G_4)-\mathcal G_5,\label{inversecomponents-2}
\\
\frac{\phi_{00}}{2\sigma^2}\epsilon^{jkl}R_{0ikl}\phi_{ij}
&=\tfrac12\mathcal G_6,\label{inversecomponents-3}
\\
\frac{1}{2\sigma^2}\epsilon^{jkl}R_{0ikl}\phi_{0i}\phi_{0j}
&=\tfrac12\mathcal G_4,\label{inversecomponents-4}
\\
\frac{1}{\sigma^2}\epsilon^{ijk}\phi_{0i}\phi_{jl}R_{0k0l}
&=\mathcal G_5,\label{inversecomponents-5}
\\
\frac{1}{\sigma^2}\epsilon^{ijk}\phi_{0i}\phi_{jl}\sum_mR_{mkml}
&=\mathcal G_3+\mathcal G_5,\label{inversecomponents-6}
\\
\frac12\epsilon^{jkl}R_{0ikl}R_{0i0j}
&=\tfrac12\mathcal G_{10},\label{inversecomponents-7}
\\
\frac12\epsilon^{jkl}R_{0ikl}\sum_mR_{mimj}
&=\tfrac12\big(\mathcal G_7+\mathcal G_{10}\big),\label{inversecomponents-8}
\\
\frac{1}{\sigma^3}\epsilon^{ijk}\phi_{0i}\phi_{jl}\nabla_{(0}\phi_{kl)}
&=-\mathcal G_8+\tfrac13\mathcal G_5,\label{inversecomponents-9}
\\
\frac{1}{2\sigma}\epsilon^{jkl}R_{0ikl}\nabla_{(0}\phi_{ij)}
&=-\tfrac12\mathcal G_9+\tfrac16\mathcal G_{10}.\label{inversecomponents-10}
\end{align}
The first six arise from curvature times two second-derivative tensors. A mixed curvature component $R_{0ijk}$ contracted with the spatial volume form can couple only to two spatial second-derivative matrices, their trace times one second-derivative tensor, $\phi_{00}$ times one second-derivative tensor, or two $\phi_{0i}$ components. Electric or purely spatial curvature requires one $\phi_{0i}$ and one spatial second-derivative matrix. These are precisely the first six lines. Curvature squared leaves only the next two lines. For two second-derivative tensors times a symmetric third derivative, the four antisymmetric slots must come from the gradient, the two second-derivative tensors, and the third derivative. Contractions connecting the two second-derivative tensors directly, or connecting both their remaining slots to the third derivative, vanish. The ninth line is the unique survivor. Curvature times the symmetric third derivative leaves only the tenth line. This proves the exhaustive upper bound used in Sec.~\ref{subsec:GSTbasis}.

Independence follows by varying the scalar and curvature components independently. Only the first two contractions have the same component degrees. At $\sigma=1$, taking $\tfrac12\epsilon^{jkl}R_{0ikl}=\operatorname{diag}(1,-1,0)$ and $\phi_{ij}=\operatorname{diag}(1,0,0)$ or $\operatorname{diag}(1,2,3)$ gives $(1,1)$ or $(-3,-6)$, with determinant $-3$. Independent electric and spatial curvature separate lines five and six, and lines seven and eight. The independent symmetric third scalar derivative separates the last two lines. The displayed map is invertible, proving independence of $\mathcal G_1,\ldots,\mathcal G_{10}$.

For the linear classification with all derivative classes included, the three additional highest-derivative contractions can be displayed in a frame parallel at the point:
\begin{equation}
\frac{1}{\sigma}\epsilon^{ijk}\phi_{il}\nabla_jR_{0k0l},\qquad
\frac{1}{\sigma}\epsilon^{ijk}\phi_{il}\sum_m\nabla_jR_{mkml},\qquad
\frac{1}{2\sigma}\phi_{0i}\epsilon^{ikl}\nabla_jR_{0jkl}.
\label{curvaturecomponents}
\end{equation}
The differential Bianchi identity removes the other divergence of the mixed curvature and expresses its normal derivative through the spatial curl of $R_{0i0j}$. The operators $(\mathcal J_1,\mathcal J_2,\mathcal J_3)$ in Eqs.~\eqref{curvaturederivatives-1}--\eqref{curvaturederivatives-3} are related to the three displayed contractions, in their order, by
\begin{equation}
\begin{pmatrix}1&-1&-1\\-1&0&0\\0&0&-2\end{pmatrix},\qquad\det=2.
\end{equation}
Independence of these three additional directions follows from the realizable ADM geometries below and the invertible correspondence \eqref{rawJ1}--\eqref{rawJ3}, avoiding a separate choice of covariant metric derivatives.

For the higher-derivative SCG contractions,
the ordering of spatial and projected Lie derivatives used in Eq.~\eqref{SCGthirdblocks} follows, for an all-lower spatial tensor, from
\begin{align}
[\mathscr L_u,\mathrm D_i]A_{j_1\cdots j_s}
&=a_i\mathscr L_u A_{j_1\cdots j_s}
-\sum_{r=1}^s C^k{}_{ij_r}A_{j_1\cdots k\cdots j_s},\notag\\
C^k{}_{ij}&=\mathrm D_iK_j{}^k+\mathrm D_jK_i{}^k-\mathrm D^kK_{ij}
+a_iK_j{}^k+a_jK_i{}^k-a^kK_{ij}.
\label{SCGmixedcomm}
\end{align}
The Ricci variation with $\partial_t h_{ij}=2NK_{ij}$, in zero-shift coordinates, gives
\begin{equation}
\mathscr L_u{}^3R_{ij}=N^{-1}\big[
\mathrm D_k\mathrm D_i(NK_j{}^k)+\mathrm D_k\mathrm D_j(NK_i{}^k)
-\mathrm D^k\mathrm D_k(NK_{ij})-\mathrm D_i\mathrm D_j(NK)\big].
\label{SCGRiccievolution}
\end{equation}
These identities introduce only products of lower-grade blocks in addition to the displayed generators. Spatial commutators give
\begin{align}
\epsilon^{ijk}K_i{}^l\mathrm D_j\mathrm D_k a_l&=-\mathcal O_1,\notag\\
\epsilon^{ijk}a_i\mathrm D^l\mathrm D_jK_{kl}&=\mathcal P_3+\mathcal O_1.
\label{SCGhighcomm}
\end{align}
They exhaust the alternative nonzero contractions of $(\mathrm D\mathrm D a)K$ and $(\mathrm D\mathrm D K)a$. For a grade-four single block, index parity and symmetry eliminate the ordered types $\mathrm D\mathscr L_u^2K$, $\mathrm D\mathscr L_u{}^3R$, $\mathrm D^3F$ and $\mathrm D\mathscr L_u^2F$. The type $\mathrm D^2\mathscr L_ua$ reduces by $\mathscr L_ua_i=\mathrm D_iF+Fa_i$. Only $\mathrm D^3K$ remains. Commuting its antisymmetrized derivatives yields the two types ${}^3R\mathrm D K$ and $(\mathrm D{}^3R)K$, including Eq.~\eqref{SCGfourthblock}.

Independence of the three additional directions can be checked on explicit local ADM geometries. Set the shift to zero and, near the origin, choose in turn
\begin{align}
&N=1,\quad h_{11}=1+2t,\quad h_{13}=h_{31}=t^2x^2,\notag\\
&N=1,\quad h_{11}=1+2t,\quad h_{13}=h_{31}=-\tfrac13(x^2)^3,\notag\\
&N=e^{x^1},\quad h_{23}=h_{32}=t(x^2)^2.
\label{SCGhighgeometries}
\end{align}
Unspecified components equal $\delta_{ij}$, and $x^2$ denotes the second spatial coordinate. At the origin all ten lower-block contractions vanish, while $(\mathcal P_1,\mathcal P_2,\mathcal P_3)$ take the values $(1,0,0)$, $(0,1,0)$, and $(0,0,1)$, respectively. These smooth local geometries satisfy the curvature identities and prove the required independence.

\subsection{Relations and independence under integrations by parts}
\label{app:independence}

Consider an identity $c_0\chi+\sum_{A=1}^7c_A\mathcal O_A=\nabla_aJ^a$ with arbitrary functions of $(t,N)$ and a local tensor current in the stated finite-order polynomial class. A current contributing at grade four can be reduced, modulo divergences of antisymmetric superpotentials, to grade three. Comparing the highest symmetric derivatives in any higher-order current removes its divergence-free principal part. Antisymmetric derivatives reduce by commutators. Repeating the step leaves a grade-three current. There is no nontrivial lower-grade parity-odd current with nonzero divergence in this class.

Its normal component is proportional to $u^a\chi$. The spatial components of the $1+1+1$ and $2+1$ partitions reduce to
\begin{equation}
u^e\varepsilon_{ebc}{}^a a^b a^dK_d{}^c,\quad
u^e\varepsilon_{ebc}{}^a \mathrm{D}^ba^dK_d{}^c,\quad
u_e\varepsilon^{eabc}K_b{}^dL_{cd},\quad
u_e\varepsilon^{eabc}K_b{}^d{}^3R_{cd},
\label{currentinventory}
\end{equation}
apart from terms involving $a\mathrm{D}K$ and $\mathrm{D}\mathrm{D}K$. The potentially nonzero $a\mathrm{D}K$ currents can be enumerated explicitly:
\begin{align}
J_1^i&=\epsilon^{ijk}a_j\mathrm{D}_kK,&J_2^i&=\epsilon^{ijk}a_j\mathrm{D}^lK_{kl},\notag\\
J_3^i&=\epsilon^{ijk}a^l\mathrm{D}_jK_{kl},&J_4^i&=\epsilon^{jkl}a_j\mathrm{D}_kK_l{}^i.
\end{align}
Antisymmetrization over four spatial indices gives $J_2+J_4=J_1+J_3$. The weighted divergence of $J_1$ vanishes, since $\mathrm{D}_ia_j$ is symmetric and $a_iJ_1^i=0$. With projected four-dimensional indices,
\begin{equation}
\nabla_b(fu_e\varepsilon^{eabc}a^dK_{cd})
=fJ_3^a+fu_e\varepsilon^{eabc}(\mathrm{D}_ba^d)K_{cd}
+(f+Nf_N)u_e\varepsilon^{eabc}a_ba^dK_{cd}.
\label{superpotential1}
\end{equation}
The left-hand side has identically zero divergence because its superpotential is antisymmetric. Thus $J_3$ and $J_4$ add no effective current beyond Eq.~\eqref{currentinventory} and $J_2$.

The remaining $\mathrm{D}\mathrm{D}K$ current, after commuting derivatives, is $J_5^i=\epsilon^{ijk}\mathrm{D}_j\mathrm{D}^lK_{kl}$. Its spatial divergence is zero, but its four-divergence includes $a_iJ_5^i$. The correct weighted superpotential relation is
\begin{equation}
\nabla_b(fu_e\varepsilon^{eabc}\mathrm{D}^dK_{cd})=fJ_5^a+(f+Nf_N)J_2^a.
\label{superpotential2}
\end{equation}
It eliminates $J_5$. The divergence of the remaining $J_2$ contains a symmetric $\mathrm{D}\mathrm{D}K$ jet absent from the proposed action representatives and from the other effective currents. Its coefficient must vanish. The $a\mathrm{D}F$ current has identically vanishing weighted divergence. Symmetric matrices such as $\mathscr L_uL_{ij}$ cannot furnish a further spatial pseudovector by themselves. These observations exhaust the grade-three partitions, including the single grade-three block.

It remains to test the five currents $u^a\chi$ and Eq.~\eqref{currentinventory}, each with its own coefficient. The divergence of $u_e\varepsilon^{eabc}K_b{}^d{}^3R_{cd}$ contains the independent $K \mathrm{D}{}^3R$ jet. Its coefficient must vanish because that jet is absent from Eq.~\eqref{PVactionbasis}. The divergences of $u^a\chi$ and $u_e\varepsilon^{eabc}K_b{}^dL_{cd}$ can cancel their $K\mathrm{D}L$ terms only with equal coefficients. Their sum then contains $2\mathcal O_{10}$, by Eq.~\eqref{weightedIBP-3} with $f=1$, so that common coefficient also vanishes. The divergence of the $\mathrm{D}aK$ current contains $\mathcal O_9$, forcing its coefficient to zero. Finally the $aaK$ current contains $\mathcal O_8$, forcing its coefficient to zero. Derivatives of the coefficients cannot cancel these highest independent jets. With the normal current removed, there is also no isolated $c_0\chi$ divergence. Linear independence of the remaining representatives then gives $c_A=0$ for all $A$.

This proves independence under local tensorial integrations by parts with general coefficients, simultaneously with linear independence. If non-tensorial characteristic currents are also admitted, the Pontryagin density
\begin{equation}
F_1^{(2;0,0)}\equiv\varepsilon_{abcd}R_{ef}{}^{cd}R^{abef}=16(\mathcal O_1-\mathcal O_5)+8\nabla_aJ_D^a
\label{Pontryagin}
\end{equation}
has no bulk variation for a constant coefficient on a fixed topological sector. This gives the additional equivalence
\begin{equation}
(c_1,c_5)\longmapsto(c_1+\kappa,c_5-\kappa),\qquad \kappa=\text{constant}.
\label{topologicalquotient}
\end{equation}
For a nonconstant coefficient, its derivative multiplies the Chern--Simons form and survives. This special equivalence is outside the tensor-current statement above.

\section{Functional variations and  Poisson brackets among primary-secondary constraints}
\label{app:variations}
For an arbitrary symmetric variation $Z_{ij}$, define
\begin{equation}
\delta_Q\chi[Z](\vec{x})\equiv
\left.\frac{\mathrm{d}}{\mathrm{d}\epsilon}\chi[Q+\epsilon Z](\vec{x})\right|_{\epsilon=0}
=\int\mathrm{d}^3y\,\frac{\delta\chi(\vec{x})}{\delta Q_{ij}(\vec{y})}Z_{ij}(\vec{y}).
\label{variationdefinition}
\end{equation}
The same notation applies to variations of $N,A,h_{ij}$. With $\chi[Q]=\epsilon^{ijk}Q_i{}^l\mathrm D_jQ_{kl}$,
\begin{equation}
\delta_Q\chi[Z]=Z_{ij}\epsilon^{(i|kl}\mathrm D_kQ_l{}^{|j)}
+Q_{ij}\epsilon^{(i|kl}\mathrm D_kZ_l{}^{|j)}.
\label{localQvariation}
\end{equation}
The weighted Euler derivative follows by one spatial integration by parts and yields the $c_1$ terms in Eq.~\eqref{beta} for weight $Nc_1A$.

For $Y_{ij}=\delta h_{ij}$ set $Y=h^{ij}Y_{ij}$ and $Y^{ij}=h^{ik}h^{jl}Y_{kl}$. In particular $\delta h^{ij}=-Y^{ij}$. Writing the spatial connection coefficients as $\varGamma^p_{ij}$, the elementary variations are
\begin{equation}
\delta\sqrt h=\tfrac12\sqrt hY,\quad
\delta\epsilon^{ijk}=-\tfrac12Y\epsilon^{ijk},\quad
\delta\varGamma^p_{ij}=\tfrac12h^{pl}(\mathrm{D}_iY_{jl}+\mathrm{D}_jY_{il}-\mathrm{D}_lY_{ij}).
\label{metricidentities}
\end{equation}
For a scalar weight $w$ held fixed under the metric variation, and at fixed lower-index $Q_{ij}$, these imply
\begin{align}
\delta_h(\sqrt h\chi)[Y]=-\sqrt h\epsilon^{ijk}\big[&Y^{lm}Q_{im}\mathrm{D}_jQ_{kl}
+\tfrac12Q_i{}^lQ_k{}^r(\mathrm{D}_jY_{lr}+\mathrm{D}_lY_{jr}-\mathrm{D}_rY_{jl})\big],\label{metriclocal}\\
\frac{\delta}{\delta h_{ij}}\int \mathrm{d}^3x\sqrt h\,w\chi
=\sqrt h\big[&-w\epsilon^{abc}Q_a{}^{(i}\mathrm{D}_bQ_c{}^{j)}
+\mathrm{D}_l(w\epsilon^{a(i|c}Q_a{}^lQ_c{}^{|j)})\big].
\label{metricEuler}
\end{align}
The connection terms in Eq.~\eqref{metriclocal} give the spatial divergence in Eq.~\eqref{metricEuler}. A local conformal variation gives
\begin{equation}
\delta_h(\sqrt h\chi)[zh]=-z\sqrt h\chi.
\label{conformalchi}
\end{equation}
For $Y_{ij}=2NQ_{ij}$, the inverse-metric variation in Eq.~\eqref{metriclocal} is $-2N\sqrt h\mathcal O_3[Q]$, whereas the two surviving connection terms sum to $2N\sqrt h\mathcal O_2[Q]$. All $\mathrm{D}_iN$ terms contract symmetric powers of $Q$ with $\epsilon$ and vanish. Equation~\eqref{cubicidentity} then proves Eq.~\eqref{metricvelocity}.

The derivatives of the unreduced $H_L$ of Eq.~\eqref{H0} are
\begin{align}
\frac{\delta H_L}{\delta N}&=\lambda_{(1)}^{ij}Q_{ij}
-\sqrt h[{}^3R+Q_{ij}Q^{ij}-Q^2+c_3+Nc_{3N}],\label{H0N}\\
\frac{\delta H_L}{\delta h_{ij}}&=\sqrt h\bigg\{
N\left[{}^3R^{ij}-\tfrac12h^{ij}{}^3R+2Q^i{}_kQ^{kj}-2QQ^{ij}
-\tfrac12h^{ij}(Q_{kl}Q^{kl}-Q^2+c_3)\right]\notag\\
&\hspace{26mm}+h^{ij}\mathrm{D}^2N-\mathrm{D}^i\mathrm{D}^jN\bigg\}.
\label{H0h}
\end{align}
The upper-index $\lambda_{(1)}^{ij}$ is an independent coordinate and has no metric variation. For $H_F=H_L+\int(N^2\lambda A-\sqrt hNc_1A\chi)\mathrm{d}^3x$,
\begin{align}
\frac{\delta H_F}{\delta N}&=\frac{\delta H_L}{\delta N}+2N\lambda A-\sqrt h A\partial_N(Nc_1)\chi,\\
\frac{\delta H_F}{\delta h_{ij}}&=\frac{\delta H_L}{\delta h_{ij}}
+\sqrt hNc_1A\epsilon^{abc}Q_a{}^{(i}\mathrm{D}_bQ_c{}^{j)}
-\sqrt h\mathrm{D}_l(Nc_1A\epsilon^{a(i|c}Q_a{}^lQ_c{}^{|j)}).
\label{HFderivatives}
\end{align}
These formulas fix $\varXi,\mu^{ij}$ in Eq.~\eqref{Fprimarypreserve} and the corresponding auxiliary multipliers in the other two models.

The remaining primary--secondary brackets are the directional derivatives below, together with Eq.~\eqref{Fmultiplieroperator}:
\begin{align}
\delta_N\alpha[W]&=[-2N\lambda+\sqrt h\partial_N(Nc_1)\chi]W,\\
\delta_h\alpha[Y]&=Nc_1\delta_h(\sqrt h\chi)[Y],\qquad
\delta_A\alpha[W]=0,\\
[\alpha(\vec{x}),\varPi(\vec{y})]&=-N^2(\vec{x})\delta^3(\vec{x}-\vec{y}),\qquad [\alpha(\vec{x}),\varPi^{(1)}_{kl}(\vec{y})]=0.
\label{alphabrackets}
\end{align}
The local metric expression is Eq.~\eqref{metriclocal}, and $\delta_Q\alpha$, $\delta_A\beta$, and $\delta_Q\beta$ are given in Eq.~\eqref{Fmultiplieroperator}. The remaining tensor derivatives are
\begin{align}
\delta_N\beta^{ij}[W]={}&-\lambda_{(1)}^{ij}W+2\sqrt h(Q^{ij}-Qh^{ij})W\notag\\
&+\sqrt h\epsilon^{(i|kl}\left[2A\partial_N(Nc_1)W \mathrm{D}_kQ_l{}^{|j)}
+\mathrm{D}_k(A\partial_N(Nc_1)W)Q_l{}^{|j)}\right],\\
\delta_h\beta^{ij}[Y]={}&2N\sqrt h\left[\tfrac12Y(Q^{ij}-Qh^{ij})-2Y^{(i}{}_kQ^{j)k}
+(Q^{kl}Y_{kl})h^{ij}+QY^{ij}\right]\notag\\
&+\sqrt h\epsilon^{(i|kl}\bigg[-2Nc_1A \mathrm{D}_k(Y^{|j)m}Q_{lm})
-\mathrm{D}_k(Nc_1A)Y^{|j)m}Q_{lm}\notag\\
&\hspace{23mm}+Nc_1Ah^{|j)r}(\mathrm{D}_kY_{pr}+\mathrm{D}_pY_{kr}-\mathrm{D}_rY_{kp})Q_l{}^p\bigg],\\
[\beta^{ij}(\vec{x}),\varPi^{(1)}_{kl}(\vec{y})]={}&-N(\vec{x})\delta^{ij}_{kl}\delta^3(\vec{x}-\vec{y}),\qquad[\beta^{ij}(\vec{x}),\varPi(\vec{y})]=0.
\label{betabrackets}
\end{align}
The spatial brackets follow from Eq.~\eqref{spatialclosure}, and the secondary--secondary brackets vanish in the unreduced phase space because $\alpha,\beta^{ij}$ depend only on coordinates. The explicit time derivatives are
\begin{align}
\partial_t\alpha&=\sqrt hNc_{1t}\chi,\\
\partial_t\beta^{ij}&=\sqrt h\epsilon^{(i|kl}
\left[2NAc_{1t}\mathrm{D}_kQ_l{}^{|j)}+\mathrm{D}_k(NAc_{1t})Q_l{}^{|j)}\right].
\label{Fexplicit}
\end{align}

\section{Regularity of the multiplier equations}
\label{app:regularity}
For the lapse-velocity interaction, a concrete constrained configuration is
\begin{equation}
h_{ij}=\delta_{ij},\quad N=N_0>0,\quad A=0,\quad
Q_{ij}=\operatorname{diag}(q_1,q_2,q_3),\quad
p=0,\quad p_{(1)}^{ij}=Q^{ij}-Q\delta^{ij},
\label{Fwitness}
\end{equation}
with $p_{(2)}^{ij}=0$ and the auxiliary momenta and multipliers set by Eq.~\eqref{Fprimary}. It satisfies all primary and secondary constraints. Let $c_1(t,N_0)\ne0$ and let the $q_i$ be pairwise different. The tensor homogeneous equations imply $Z_{ii}=0$ and
\begin{equation}
Z_{12}=-\frac{c_1}{4}(q_1-q_2)\partial_3W,\quad
Z_{23}=-\frac{c_1}{4}(q_2-q_3)\partial_1W,\quad
Z_{31}=-\frac{c_1}{4}(q_3-q_1)\partial_2W.
\end{equation}
The remaining scalar equation is
\begin{equation}
\frac{N_0c_1^2}{4}\left[(q_2-q_3)^2\partial_1^2
+(q_3-q_1)^2\partial_2^2+(q_1-q_2)^2\partial_3^2\right]W=0.
\label{Felliptic}
\end{equation}
For pairwise distinct $q_i$ and $c_1\ne0$, homogeneous Dirichlet data on a bounded smooth domain force $W=0$, and hence $Z_{ij}=0$. With smooth sources, the same elimination gives an inhomogeneous elliptic Dirichlet problem and determines all seven multipliers. The scalar Dirichlet condition also removes the adjoint boundary pairing. On a periodic domain, the constant scalar mode instead requires a source compatibility condition.

The general operator includes derivatives proportional to $A$ and can develop additional kernels. For constant $A=A_0$ and a wave vector along the third principal axis, the $(Z_{13},Z_{23})$ tensor block has determinant $N_0^2(4-c_1^2A_0^2k_3^2)$. Resonant modes must be excluded or treated by separate compatibility conditions. The count \eqref{Fcount} therefore applies to regular operator domains.

Eliminating the 14 auxiliary constraints sets $\lambda=p$ and $\lambda_{(1)}^{ij}=2p_{(1)}^{ij}$ without changing the remaining canonical brackets. The nonspatial constraint matrix then has blocks $\bigl(\begin{smallmatrix}0&-J^*\\J&K\end{smallmatrix}\bigr)$, where $J$ is Eq.~\eqref{Fmultiplieroperator}, $J^*$ its adjoint, and $K$ the reduced secondary--secondary bracket. Invertibility of $J$ and $J^*$ makes this matrix invertible independently of $K$, proving the second-class count used in Sec.~\ref{sec:F}.

For $c_2a\ne0$, the tensor-primary bracket in Eq.~\eqref{Lprimaryactions} is the commutator of a symmetric tensor with the rotation generator about $a_i$. Its rank is four, with kernel spanned by $h_{ij}$ and $a_i a_j-a^2h_{ij}/3$, where $a^2=h^{ij}a_i a_j$. Define
\begin{equation}
v^r=\epsilon^r{}_{mn}Q^{mi}a_i a^n,
\qquad v^r a_r=0.
\label{vdef}
\end{equation}

We call the trace-free kernel tensor $a_i a_j-a^2h_{ij}/3$ the axial direction. Write
\begin{equation}
\nu^{(2)}_{ij}=\nu^\perp_{ij}+z\left(a_i a_j-\frac{a^2}{3}h_{ij}\right)+z_0h_{ij},
\qquad h^{ij}\nu^\perp_{ij}=a^ia^j\nu^\perp_{ij}=0.
\label{multiplierdecomposition}
\end{equation}

Define also
\begin{equation}
U=a_i a_j\epsilon^{(i|kl}\mathrm D_kQ_l{}^{|j)}.
\label{Udef}
\end{equation}
For the interactions containing $L_{ij}$, let $W$ and $Z_{ij}$ denote homogeneous variations of $\xi$ and $\nu^{(2)}_{ij}$, with $Z_{ij}$ decomposed as in Eq.~\eqref{multiplierdecomposition}. The single-interaction case follows by setting $c_1=0$.
For the axial tensor $Z_{ij}=z(a_i a_j-a^2h_{ij}/3)$, direct product differentiation gives
\begin{align}
Z_{ij}\epsilon^{(i|kl}\mathrm{D}_kQ_l{}^{|j)}&=Uz,\\
Q_{ij}\epsilon^{(i|kl}\mathrm{D}_kZ_l{}^{|j)}&=Uz-\mathrm{D}_r(v^rz).
\label{axialidentities}
\end{align}
No derivative of $a_i$ is omitted in these equalities. Moreover $v^r\mathrm{D}_r[(c_2-c_1)/N]=0$, since $v^ra_r=0$ and the coefficients depend on $t,N$. The resulting scalar and tensor axial operators have opposite transpose signs, as required by bracket antisymmetry.

The homogeneous axial projection of Eq.~\eqref{bothtensor} is
\begin{equation}
(c_2-c_1)v^r\mathrm{D}_rW-2c_1UW=0.
\label{axialhomogeneous}
\end{equation}
Once $W=0$ is fixed, the four nonkernel tensor equations give $Z^\perp_{ij}=0$. The scalar equation then reduces to
\begin{equation}
(c_2-c_1)\mathrm{D}_r(v^rz)+2c_1Uz=0.
\label{bothzHom}
\end{equation}
The mixed brackets have the boundary pairing
\begin{equation}
\int_{\partial\varOmega}\mathrm{d}\varSigma\,\frac{c_1-c_2}{N}
W n_r\epsilon^r{}_{mn}Q^{mi}Z_i{}^n,
\label{bothboundarypair}
\end{equation}
where $\mathrm{d}\varSigma$ is the induced area element and $n_r$ the outward unit normal. For $c_2-c_1\ne0$ and $v\ne0$ on a smooth open flow tube, impose the dual homogeneous boundary conditions
\begin{equation}
W|_{\varGamma_{\rm in}}=0,\qquad
n_r\epsilon^r{}_{mn}Q^{mi}Z_i{}^n|_{\varGamma_{\rm out}}=0,
\label{dualboundary}
\end{equation}
with transverse compact support. Here $\varGamma_{\rm in}$ and $\varGamma_{\rm out}$ are the inflow and outflow surfaces of $v$. These data remove the boundary pairing: the first equation fixes $W$, the four nonkernel equations fix $Z^\perp$, and the opposite-endpoint condition fixes $z$. The same sequence solves the sourced equations for smooth compatible data. Closed characteristics can instead admit zero modes and impose compatibility conditions.

A constrained configuration realizing this transport case is $h_{ij}=\delta_{ij}$, $N=e^{x^3}$, $Q_{13}=Q_{31}=q\ne0$, and all other $Q_{ij}=0$, with $p=0$, $p_{(1)}^{ij}=Q^{ij}$, and $p_{(2)}^{ij}=B^{ij}$. It gives $v=-q\partial_2$, $\chi=U=0$, and $S=\mathcal C_i=0$, with the auxiliary constraints imposed. It applies to both the single-interaction theory and the combined theory with $c_1\ne c_2$.

For $c_1=c_2=c\ne0$, both axial equations are algebraic with nonzero coefficients when $U\ne0$. An explicit constrained configuration has $h_{ij}=\delta_{ij}$, $N=e^{x^3}$, $Q_{13}=Q_{31}=x^2$, $p=0$, $p_{(1)}^{12}=p_{(1)}^{21}=-c(x^2)^2/2$, and $p_{(2)}^{23}=p_{(2)}^{32}=c x^2/2$, with other components zero and the auxiliary constraints imposed. Here $x^2$ is the second spatial coordinate. It gives $Q=\chi=0$, $U=-1$, $v=-x^2\partial_2$, and $S=\mathcal C_i=0$. The coefficients depend on $t,N$ and are constant along $x^2$, so the spatial constraints hold also for general $c(t,N)$. This verifies that the algebraic case occurs on the constraint surface, including where $v=0$.

The two examples show that neither $c_1=c_2\ne0$ nor $c_1=-c_2\ne0$ creates an additional null direction for general configurations. On these regular domains, the four nonkernel tensor components, the two scalar equations \eqref{axialhomogeneous} and \eqref{bothzHom}, and the trace condition \eqref{bothtracefix} close the multiplier system. For $c_1=0$, use Eq.~\eqref{tracefixL} instead. This establishes the rank assumptions used in Secs.~\ref{sec:L} and \ref{sec:both}, without extending them to boundary zero modes or other kinetic completions.

%%%

\providecommand{\href}[2]{#2}\begingroup\raggedright\endgroup


\begin{thebibliography}{10}
	
	\bibitem{Nicolis:2008in}
	A.~Nicolis, R.~Rattazzi and E.~Trincherini, \emph{{The Galileon as a local
			modification of gravity}},
	\href{https://doi.org/10.1103/PhysRevD.79.064036}{\emph{Phys.Rev.} {\bfseries
			D79} (2009) 064036} [\href{https://arxiv.org/abs/0811.2197}{{\ttfamily
			0811.2197}}].
	
	\bibitem{Deffayet:2009wt}
	C.~Deffayet, G.~Esposito-Farese and A.~Vikman, \emph{{Covariant Galileon}},
	\href{https://doi.org/10.1103/PhysRevD.79.084003}{\emph{Phys.Rev.} {\bfseries
			D79} (2009) 084003} [\href{https://arxiv.org/abs/0901.1314}{{\ttfamily
			0901.1314}}].
	
	\bibitem{Horndeski:1974wa}
	G.W.~Horndeski, \emph{{Second-order scalar-tensor field equations in a
			four-dimensional space}},
	\href{https://doi.org/10.1007/BF01807638}{\emph{Int.J.Theor.Phys.} {\bfseries
			10} (1974) 363}.
	
	\bibitem{Deffayet:2011gz}
	C.~Deffayet, X.~Gao, D.~Steer and G.~Zahariade, \emph{{From k-essence to
			generalised Galileons}},
	\href{https://doi.org/10.1103/PhysRevD.84.064039}{\emph{Phys.Rev.} {\bfseries
			D84} (2011) 064039} [\href{https://arxiv.org/abs/1103.3260}{{\ttfamily
			1103.3260}}].
	
	\bibitem{Kobayashi:2011nu}
	T.~Kobayashi, M.~Yamaguchi and J.~Yokoyama, \emph{{Generalized G-inflation:
			Inflation with the most general second-order field equations}},
	\href{https://doi.org/10.1143/PTP.126.511}{\emph{Prog.Theor.Phys.} {\bfseries
			126} (2011) 511} [\href{https://arxiv.org/abs/1105.5723}{{\ttfamily
			1105.5723}}].
	
	\bibitem{Gleyzes:2014dya}
	J.~Gleyzes, D.~Langlois, F.~Piazza and F.~Vernizzi, \emph{{Healthy theories
			beyond Horndeski}},
	\href{https://doi.org/10.1103/PhysRevLett.114.211101}{\emph{Phys. Rev. Lett.}
		{\bfseries 114} (2015) 211101}
	[\href{https://arxiv.org/abs/1404.6495}{{\ttfamily 1404.6495}}].
	
	\bibitem{Langlois:2015cwa}
	D.~Langlois and K.~Noui, \emph{{Degenerate higher derivative theories beyond
			Horndeski: evading the Ostrogradski instability}},
	\href{https://doi.org/10.1088/1475-7516/2016/02/034}{\emph{JCAP} {\bfseries
			1602} (2016) 034} [\href{https://arxiv.org/abs/1510.06930}{{\ttfamily
			1510.06930}}].
	
	\bibitem{BenAchour:2016fzp}
	J.~Ben~Achour, M.~Crisostomi, K.~Koyama, D.~Langlois, K.~Noui and G.~Tasinato,
	\emph{{Degenerate higher order scalar-tensor theories beyond Horndeski up to
			cubic order}}, \href{https://doi.org/10.1007/JHEP12(2016)100}{\emph{JHEP}
		{\bfseries 12} (2016) 100}
	[\href{https://arxiv.org/abs/1608.08135}{{\ttfamily 1608.08135}}].
	
	\bibitem{Langlois:2015skt}
	D.~Langlois and K.~Noui, \emph{{Hamiltonian analysis of higher derivative
			scalar-tensor theories}},
	\href{https://doi.org/10.1088/1475-7516/2016/07/016}{\emph{JCAP} {\bfseries
			1607} (2016) 016} [\href{https://arxiv.org/abs/1512.06820}{{\ttfamily
			1512.06820}}].
	
	\bibitem{deRham:2010eu}
	C.~de~Rham and A.J.~Tolley, \emph{{DBI and the Galileon reunited}},
	\href{https://doi.org/10.1088/1475-7516/2010/05/015}{\emph{JCAP} {\bfseries
			1005} (2010) 015} [\href{https://arxiv.org/abs/1003.5917}{{\ttfamily
			1003.5917}}].
	
	\bibitem{Goon:2011qf}
	G.~Goon, K.~Hinterbichler and M.~Trodden, \emph{{Symmetries for Galileons and
			DBI scalars on curved space}},
	\href{https://doi.org/10.1088/1475-7516/2011/07/017}{\emph{JCAP} {\bfseries
			1107} (2011) 017} [\href{https://arxiv.org/abs/1103.5745}{{\ttfamily
			1103.5745}}].
	
	\bibitem{Bekenstein:1992pj}
	J.D.~Bekenstein, \emph{{The Relation between physical and gravitational
			geometry}}, \href{https://doi.org/10.1103/PhysRevD.48.3641}{\emph{Phys. Rev.}
		{\bfseries D48} (1993) 3641}
	[\href{https://arxiv.org/abs/gr-qc/9211017}{{\ttfamily gr-qc/9211017}}].
	
	\bibitem{Bettoni:2013diz}
	D.~Bettoni and S.~Liberati, \emph{{Disformal invariance of second order
			scalar-tensor theories: Framing the Horndeski action}},
	\href{https://doi.org/10.1103/PhysRevD.88.084020}{\emph{Phys.Rev.} {\bfseries
			D88} (2013) 084020} [\href{https://arxiv.org/abs/1306.6724}{{\ttfamily
			1306.6724}}].
	
	\bibitem{Zumalacarregui:2013pma}
	M.~Zumalacárregui and J.~García-Bellido, \emph{{Transforming gravity: from
			derivative couplings to matter to second-order scalar-tensor theories beyond
			the Horndeski Lagrangian}},
	\href{https://doi.org/10.1103/PhysRevD.89.064046}{\emph{Phys.Rev.} {\bfseries
			D89} (2014) 064046} [\href{https://arxiv.org/abs/1308.4685}{{\ttfamily
			1308.4685}}].
	
	\bibitem{BenAchour:2016cay}
	J.~Ben~Achour, D.~Langlois and K.~Noui, \emph{{Degenerate higher order
			scalar-tensor theories beyond Horndeski and disformal transformations}},
	\href{https://doi.org/10.1103/PhysRevD.93.124005}{\emph{Phys. Rev. D}
		{\bfseries 93} (2016) 124005}
	[\href{https://arxiv.org/abs/1602.08398}{{\ttfamily 1602.08398}}].
	
	\bibitem{Takahashi:2017zgr}
	K.~Takahashi, H.~Motohashi, T.~Suyama and T.~Kobayashi, \emph{{General
			invertible transformation and physical degrees of freedom}},
	\href{https://doi.org/10.1103/PhysRevD.95.084053}{\emph{Phys. Rev.}
		{\bfseries D95} (2017) 084053}
	[\href{https://arxiv.org/abs/1702.01849}{{\ttfamily 1702.01849}}].
	
	\bibitem{Gao:2014soa}
	X.~Gao, \emph{{Unifying framework for scalar-tensor theories of gravity}},
	\href{https://doi.org/10.1103/PhysRevD.90.081501}{\emph{Phys.Rev.} {\bfseries
			D90} (2014) 081501} [\href{https://arxiv.org/abs/1406.0822}{{\ttfamily
			1406.0822}}].
	
	\bibitem{Gao:2014fra}
	X.~Gao, \emph{{Hamiltonian analysis of spatially covariant gravity}},
	\href{https://doi.org/10.1103/PhysRevD.90.104033}{\emph{Phys.Rev.} {\bfseries
			D90} (2014) 104033} [\href{https://arxiv.org/abs/1409.6708}{{\ttfamily
			1409.6708}}].
	
	\bibitem{Gao:2018znj}
	X.~Gao and Z.-B.~Yao, \emph{{Spatially covariant gravity with velocity of the
			lapse function: the Hamiltonian analysis}},
	\href{https://doi.org/10.1088/1475-7516/2019/05/024}{\emph{JCAP} {\bfseries
			1905} (2019) 024} [\href{https://arxiv.org/abs/1806.02811}{{\ttfamily
			1806.02811}}].
	
	\bibitem{Gao:2019lpz}
	X.~Gao, C.~Kang and Z.-B.~Yao, \emph{{Spatially Covariant Gravity: Perturbative
			Analysis and Field Transformations}},
	\href{https://doi.org/10.1103/PhysRevD.99.104015}{\emph{Phys. Rev.}
		{\bfseries D99} (2019) 104015}
	[\href{https://arxiv.org/abs/1902.07702}{{\ttfamily 1902.07702}}].
	
	\bibitem{Cheung:2007st}
	C.~Cheung, P.~Creminelli, A.L.~Fitzpatrick, J.~Kaplan and L.~Senatore,
	\emph{{The Effective Field Theory of Inflation}},
	\href{https://doi.org/10.1088/1126-6708/2008/03/014}{\emph{JHEP} {\bfseries
			0803} (2008) 014} [\href{https://arxiv.org/abs/0709.0293}{{\ttfamily
			0709.0293}}].
	
	\bibitem{Gubitosi:2012hu}
	G.~Gubitosi, F.~Piazza and F.~Vernizzi, \emph{{The Effective Field Theory of
			Dark Energy}},
	\href{https://doi.org/10.1088/1475-7516/2013/02/032}{\emph{JCAP} {\bfseries
			1302} (2013) 032} [\href{https://arxiv.org/abs/1210.0201}{{\ttfamily
			1210.0201}}].
	
	\bibitem{Gleyzes:2014rba}
	J.~Gleyzes, D.~Langlois and F.~Vernizzi, \emph{{A unifying description of dark
			energy}}, \href{https://doi.org/10.1142/S021827181443010X}{\emph{Int. J. Mod.
			Phys.} {\bfseries D23} (2014) 1443010}
	[\href{https://arxiv.org/abs/1411.3712}{{\ttfamily 1411.3712}}].
	
	\bibitem{Horava:2009uw}
	P.~Horava, \emph{{Quantum Gravity at a Lifshitz Point}},
	\href{https://doi.org/10.1103/PhysRevD.79.084008}{\emph{Phys.Rev.} {\bfseries
			D79} (2009) 084008} [\href{https://arxiv.org/abs/0901.3775}{{\ttfamily
			0901.3775}}].
	
	\bibitem{Blas:2009qj}
	D.~Blas, O.~Pujolas and S.~Sibiryakov, \emph{{Consistent Extension of Horava
			Gravity}},
	\href{https://doi.org/10.1103/PhysRevLett.104.181302}{\emph{Phys.Rev.Lett.}
		{\bfseries 104} (2010) 181302}
	[\href{https://arxiv.org/abs/0909.3525}{{\ttfamily 0909.3525}}].
	
	\bibitem{deRham:2010kj}
	C.~de~Rham, G.~Gabadadze and A.J.~Tolley, \emph{{Resummation of Massive
			Gravity}},
	\href{https://doi.org/10.1103/PhysRevLett.106.231101}{\emph{Phys.Rev.Lett.}
		{\bfseries 106} (2011) 231101}
	[\href{https://arxiv.org/abs/1011.1232}{{\ttfamily 1011.1232}}].
	
	\bibitem{Endlich:2012pz}
	S.~Endlich, A.~Nicolis and J.~Wang, \emph{{Solid Inflation}},
	\href{https://doi.org/10.1088/1475-7516/2013/10/011}{\emph{JCAP} {\bfseries
			1310} (2013) 011} [\href{https://arxiv.org/abs/1210.0569}{{\ttfamily
			1210.0569}}].
	
	\bibitem{Gao:2020juc}
	X.~Gao, \emph{{Higher derivative scalar-tensor monomials and their
			classification}}, \href{https://doi.org/10.1007/s11433-020-1607-3}{\emph{Sci.
			China Phys. Mech. Astron.} {\bfseries 64} (2021) 210012}
	[\href{https://arxiv.org/abs/2003.11978}{{\ttfamily 2003.11978}}].
	
	\bibitem{Gao:2020yzr}
	X.~Gao and Y.-M.~Hu, \emph{{Higher derivative scalar-tensor theory and
			spatially covariant gravity: the correspondence}},
	\href{https://doi.org/10.1103/PhysRevD.102.084006}{\emph{Phys. Rev. D}
		{\bfseries 102} (2020) 084006}
	[\href{https://arxiv.org/abs/2004.07752}{{\ttfamily 2004.07752}}].
	
	\bibitem{Gao:2020qxy}
	X.~Gao, \emph{{Higher derivative scalar-tensor theory from the spatially
			covariant gravity: a linear algebraic analysis}},
	\href{https://doi.org/10.1088/1475-7516/2020/11/004}{\emph{JCAP} {\bfseries
			11} (2020) 004} [\href{https://arxiv.org/abs/2006.15633}{{\ttfamily
			2006.15633}}].
	
	\bibitem{Hu:2021bbo}
	Y.-M.~Hu and X.~Gao, \emph{{Covariant 3+1 correspondence of the spatially
			covariant gravity and the degeneracy conditions}},
	\href{https://doi.org/10.1103/PhysRevD.105.044023}{\emph{Phys. Rev. D}
		{\bfseries 105} (2022) 044023}
	[\href{https://arxiv.org/abs/2111.08652}{{\ttfamily 2111.08652}}].
	
	\bibitem{DeFelice:2018ewo}
	A.~De~Felice, D.~Langlois, S.~Mukohyama, K.~Noui and A.~Wang,
	\emph{{Generalized instantaneous modes in higher-order scalar-tensor
			theories}}, \href{https://doi.org/10.1103/PhysRevD.98.084024}{\emph{Phys.
			Rev. D} {\bfseries 98} (2018) 084024}
	[\href{https://arxiv.org/abs/1803.06241}{{\ttfamily 1803.06241}}].
	
	\bibitem{DeFelice:2021hps}
	A.~De~Felice, S.~Mukohyama and K.~Takahashi, \emph{{Nonlinear definition of the
			shadowy mode in higher-order scalar-tensor theories}},
	\href{https://doi.org/10.1088/1475-7516/2021/12/020}{\emph{JCAP} {\bfseries
			12} (2021) 020} [\href{https://arxiv.org/abs/2110.03194}{{\ttfamily
			2110.03194}}].
	
	\bibitem{Gao:2019twq}
	X.~Gao and Z.-B.~Yao, \emph{{Spatially covariant gravity theories with two
			tensorial degrees of freedom: the formalism}},
	\href{https://doi.org/10.1103/PhysRevD.101.064018}{\emph{Phys. Rev.}
		{\bfseries D101} (2020) 064018}
	[\href{https://arxiv.org/abs/1910.13995}{{\ttfamily 1910.13995}}].
	
	\bibitem{Hu:2021yaq}
	Y.-M.~Hu and X.~Gao, \emph{{Spatially covariant gravity with 2 degrees of
			freedom: Perturbative analysis}},
	\href{https://doi.org/10.1103/PhysRevD.104.104007}{\emph{Phys. Rev. D}
		{\bfseries 104} (2021) 104007}
	[\href{https://arxiv.org/abs/2104.07615}{{\ttfamily 2104.07615}}].
	
	\bibitem{Yu:2026fgn}
	Y.~Yu, Y.-M.~Hu and X.~Gao, \emph{{Spatially covariant gravity with two degrees
			of freedom: A perturbative analysis up to cubic order}},
	\href{https://arxiv.org/abs/2604.14490}{{\ttfamily 2604.14490}}.
	
	\bibitem{Yao:2020tur}
	Z.-B.~Yao, M.~Oliosi, X.~Gao and S.~Mukohyama, \emph{{Minimally modified
			gravity with an auxiliary constraint: A Hamiltonian construction}},
	\href{https://doi.org/10.1103/PhysRevD.103.024032}{\emph{Phys. Rev. D}
		{\bfseries 103} (2021) 024032}
	[\href{https://arxiv.org/abs/2011.00805}{{\ttfamily 2011.00805}}].
	
	\bibitem{Yao:2023qjd}
	Z.-B.~Yao, M.~Oliosi, X.~Gao and S.~Mukohyama, \emph{{Minimally modified
			gravity with auxiliary constraints formalism}},
	\href{https://doi.org/10.1103/PhysRevD.107.104052}{\emph{Phys. Rev. D}
		{\bfseries 107} (2023) 104052}
	[\href{https://arxiv.org/abs/2302.02090}{{\ttfamily 2302.02090}}].
	
	\bibitem{Gao:2018izs}
	X.~Gao, M.~Yamaguchi and D.~Yoshida, \emph{{Higher derivative scalar-tensor
			theory through a non-dynamical scalar field}},
	\href{https://doi.org/10.1088/1475-7516/2019/03/006}{\emph{JCAP} {\bfseries
			1903} (2019) 006} [\href{https://arxiv.org/abs/1810.07434}{{\ttfamily
			1810.07434}}].
	
	\bibitem{Wang:2024hfd}
	Z.-C.~Wang and X.~Gao, \emph{{Spatial covariant gravity with two degrees of
			freedom in the presence of an auxiliary scalar field: Perturbation
			analysis}}, \href{https://doi.org/10.1088/1674-1137/ad47a9}{\emph{Chin. Phys.
			C} {\bfseries 48} (2024) 085101}
	[\href{https://arxiv.org/abs/2403.15355}{{\ttfamily 2403.15355}}].
	
	\bibitem{Zhu:2025bcu}
	J.-C.~Zhu, S.-Y.~Li and X.~Gao, \emph{{Spatially covariant gravity with two
			degrees of freedom in the presence of an auxiliary scalar field: Hamiltonian
			analysis}},  \href{https://arxiv.org/abs/2508.02466}{{\ttfamily 2508.02466}}.
	
	\bibitem{Yu:2024drx}
	Y.~Yu, Z.~Chen and X.~Gao, \emph{{Spatially covariant gravity with
			nonmetricity}},
	\href{https://doi.org/10.1140/epjc/s10052-024-12893-5}{\emph{Eur. Phys. J. C}
		{\bfseries 84} (2024) 549}
	[\href{https://arxiv.org/abs/2402.02565}{{\ttfamily 2402.02565}}].
	
	\bibitem{Yu:2024sed}
	Y.~Yu, Z.~Chen, Y.-M.~Hu and X.~Gao, \emph{{Interacting hypersurfaces and
			multiple scalar-tensor theories}},
	\href{https://doi.org/10.1103/PhysRevD.111.024052}{\emph{Phys. Rev. D}
		{\bfseries 111} (2025) 024052}
	[\href{https://arxiv.org/abs/2410.12680}{{\ttfamily 2410.12680}}].
	
	\bibitem{Chen:2026hhl}
	J.-J.~Chen, X.~Gao and T.~Kobayashi, \emph{{Bi-scalar-tensor theory: a
			spatially covariant gravity approach}},
	\href{https://arxiv.org/abs/2609.23704}{{\ttfamily 2609.23704}}.
	
	\bibitem{Li:2025qbv}
	S.-Y.~Li and X.~Gao, \emph{{On the degrees of freedom of spatially covariant
			vector field theory}},  \href{https://arxiv.org/abs/2510.26639}{{\ttfamily
			2510.26639}}.
	
	\bibitem{Fujita:2015ymn}
	T.~Fujita, X.~Gao and J.~Yokoyama, \emph{{Spatially covariant theories of
			gravity: disformal transformation, cosmological perturbations and the
			Einstein frame}},
	\href{https://doi.org/10.1088/1475-7516/2016/02/014}{\emph{JCAP} {\bfseries
			1602} (2016) 014} [\href{https://arxiv.org/abs/1511.04324}{{\ttfamily
			1511.04324}}].
	
	\bibitem{Zhu:2022caa}
	X.-Z.~Zhu, Y.~Yu and X.~Gao, \emph{{Cosmological perturbations in the spatially
			covariant gravity with a dynamical lapse function}},
	\href{https://doi.org/10.1016/j.physletb.2022.137356}{\emph{Phys. Lett. B}
		{\bfseries 833} (2022) 137356}
	[\href{https://arxiv.org/abs/2208.03629}{{\ttfamily 2208.03629}}].
	
	\bibitem{Gao:2019liu}
	X.~Gao and X.-Y.~Hong, \emph{{Propagation of gravitational waves in a
			cosmological background}},
	\href{https://doi.org/10.1103/PhysRevD.101.064057}{\emph{Phys.\ Rev.\ D}
		{\bfseries 101} (2020) 064057}
	[\href{https://arxiv.org/abs/1906.07131}{{\ttfamily 1906.07131}}].
	
	\bibitem{Zhu:2022dfq}
	T.~Zhu, W.~Zhao and A.~Wang, \emph{{Polarized primordial gravitational waves in
			spatial covariant gravities}},
	\href{https://doi.org/10.1103/PhysRevD.107.024031}{\emph{Phys. Rev. D}
		{\bfseries 107} (2023) 024031}
	[\href{https://arxiv.org/abs/2210.05259}{{\ttfamily 2210.05259}}].
	
	\bibitem{Zhu:2022uoq}
	T.~Zhu, W.~Zhao and A.~Wang, \emph{{Gravitational wave constraints on spatial
			covariant gravities}},
	\href{https://doi.org/10.1103/PhysRevD.107.044051}{\emph{Phys. Rev. D}
		{\bfseries 107} (2023) 044051}
	[\href{https://arxiv.org/abs/2211.04711}{{\ttfamily 2211.04711}}].
	
	\bibitem{Mukohyama:2022skk}
	S.~Mukohyama, K.~Takahashi and V.~Yingcharoenrat, \emph{{Generalized
			Regge-Wheeler equation from Effective Field Theory of black hole
			perturbations with a timelike scalar profile}},
	\href{https://doi.org/10.1088/1475-7516/2022/10/050}{\emph{JCAP} {\bfseries
			10} (2022) 050} [\href{https://arxiv.org/abs/2208.02943}{{\ttfamily
			2208.02943}}].
	
	\bibitem{Barura:2024uog}
	C.G.A.~Barura, H.~Kobayashi, S.~Mukohyama, N.~Oshita, K.~Takahashi and
	V.~Yingcharoenrat, \emph{{Tidal Love numbers from EFT of black hole
			perturbations with timelike scalar profile}},
	\href{https://doi.org/10.1088/1475-7516/2024/09/001}{\emph{JCAP} {\bfseries
			09} (2024) 001} [\href{https://arxiv.org/abs/2405.10813}{{\ttfamily
			2405.10813}}].
	
	\bibitem{Jackiw:2003pm}
	R.~Jackiw and S.Y.~Pi, \emph{{Chern-Simons modification of general
			relativity}}, \href{https://doi.org/10.1103/PhysRevD.68.104012}{\emph{Phys.
			Rev.} {\bfseries D68} (2003) 104012}
	[\href{https://arxiv.org/abs/gr-qc/0308071}{{\ttfamily gr-qc/0308071}}].
	
	\bibitem{Satoh:2007gn}
	M.~Satoh, S.~Kanno and J.~Soda, \emph{{Circular Polarization of Primordial
			Gravitational Waves in String-inspired Inflationary Cosmology}},
	\href{https://doi.org/10.1103/PhysRevD.77.023526}{\emph{Phys. Rev.}
		{\bfseries D77} (2008) 023526}
	[\href{https://arxiv.org/abs/0706.3585}{{\ttfamily 0706.3585}}].
	
	\bibitem{Satoh:2010ep}
	M.~Satoh, \emph{{Slow-roll Inflation with the Gauss-Bonnet and Chern-Simons
			Corrections}},
	\href{https://doi.org/10.1088/1475-7516/2010/11/024}{\emph{JCAP} {\bfseries
			11} (2010) 024} [\href{https://arxiv.org/abs/1008.2724}{{\ttfamily
			1008.2724}}].
	
	\bibitem{Qiao:2019hkz}
	J.~Qiao, T.~Zhu, W.~Zhao and A.~Wang, \emph{{Polarized primordial gravitational
			waves in the ghost-free parity-violating gravity}},
	\href{https://doi.org/10.1103/PhysRevD.101.043528}{\emph{Phys. Rev. D}
		{\bfseries 101} (2020) 043528}
	[\href{https://arxiv.org/abs/1911.01580}{{\ttfamily 1911.01580}}].
	
	\bibitem{Creminelli:2014wna}
	P.~Creminelli, J.~Gleyzes, J.~Noreña and F.~Vernizzi, \emph{{Resilience of the
			standard predictions for primordial tensor modes}},
	\href{https://doi.org/10.1103/PhysRevLett.113.231301}{\emph{Phys. Rev. Lett.}
		{\bfseries 113} (2014) 231301}
	[\href{https://arxiv.org/abs/1407.8439}{{\ttfamily 1407.8439}}].
	
	\bibitem{Cannone:2015rra}
	D.~Cannone, J.-O.~Gong and G.~Tasinato, \emph{{Breaking discrete symmetries in
			the effective field theory of inflation}},
	\href{https://doi.org/10.1088/1475-7516/2015/08/003}{\emph{JCAP} {\bfseries
			1508} (2015) 003} [\href{https://arxiv.org/abs/1505.05773}{{\ttfamily
			1505.05773}}].
	
	\bibitem{Lue:1998mq}
	A.~Lue, L.-M.~Wang and M.~Kamionkowski, \emph{{Cosmological signature of new
			parity violating interactions}},
	\href{https://doi.org/10.1103/PhysRevLett.83.1506}{\emph{Phys. Rev. Lett.}
		{\bfseries 83} (1999) 1506}
	[\href{https://arxiv.org/abs/astro-ph/9812088}{{\ttfamily
			astro-ph/9812088}}].
	
	\bibitem{Saito:2007kt}
	S.~Saito, K.~Ichiki and A.~Taruya, \emph{{Probing polarization states of
			primordial gravitational waves with CMB anisotropies}},
	\href{https://doi.org/10.1088/1475-7516/2007/09/002}{\emph{JCAP} {\bfseries
			09} (2007) 002} [\href{https://arxiv.org/abs/0705.3701}{{\ttfamily
			0705.3701}}].
	
	\bibitem{Contaldi:2008yz}
	C.R.~Contaldi, J.~Magueijo and L.~Smolin, \emph{{Anomalous CMB polarization and
			gravitational chirality}},
	\href{https://doi.org/10.1103/PhysRevLett.101.141101}{\emph{Phys. Rev. Lett.}
		{\bfseries 101} (2008) 141101}
	[\href{https://arxiv.org/abs/0806.3082}{{\ttfamily 0806.3082}}].
	
	\bibitem{Gluscevic:2010vv}
	V.~Gluscevic and M.~Kamionkowski, \emph{{Testing Parity-Violating Mechanisms
			with Cosmic Microwave Background Experiments}},
	\href{https://doi.org/10.1103/PhysRevD.81.123529}{\emph{Phys. Rev.}
		{\bfseries D81} (2010) 123529}
	[\href{https://arxiv.org/abs/1002.1308}{{\ttfamily 1002.1308}}].
	
	\bibitem{LIGOScientific:2016aoc}
	{\scshape LIGO Scientific, Virgo} collaboration, \emph{{Observation of
			Gravitational Waves from a Binary Black Hole Merger}},
	\href{https://doi.org/10.1103/PhysRevLett.116.061102}{\emph{Phys. Rev. Lett.}
		{\bfseries 116} (2016) 061102}
	[\href{https://arxiv.org/abs/1602.03837}{{\ttfamily 1602.03837}}].
	
	\bibitem{LIGOScientific:2017vwq}
	{\scshape LIGO Scientific, Virgo} collaboration, \emph{{GW170817: Observation
			of Gravitational Waves from a Binary Neutron Star Inspiral}},
	\href{https://doi.org/10.1103/PhysRevLett.119.161101}{\emph{Phys. Rev. Lett.}
		{\bfseries 119} (2017) 161101}
	[\href{https://arxiv.org/abs/1710.05832}{{\ttfamily 1710.05832}}].
	
	\bibitem{Alexander:2017jmt}
	S.H.~Alexander and N.~Yunes, \emph{{Gravitational wave probes of parity
			violation in compact binary coalescences}},
	\href{https://doi.org/10.1103/PhysRevD.97.064033}{\emph{Phys. Rev.}
		{\bfseries D97} (2018) 064033}
	[\href{https://arxiv.org/abs/1712.01853}{{\ttfamily 1712.01853}}].
	
	\bibitem{Qiao:2019wsh}
	J.~Qiao, T.~Zhu, W.~Zhao and A.~Wang, \emph{{Waveform of gravitational waves in
			the ghost-free parity-violating gravities}},
	\href{https://doi.org/10.1103/PhysRevD.100.124058}{\emph{Phys. Rev.}
		{\bfseries D100} (2019) 124058}
	[\href{https://arxiv.org/abs/1909.03815}{{\ttfamily 1909.03815}}].
	
	\bibitem{Zhao:2019xmm}
	W.~Zhao, T.~Zhu, J.~Qiao and A.~Wang, \emph{{Waveform of gravitational waves in
			the general parity-violating gravities}},
	\href{https://doi.org/10.1103/PhysRevD.101.024002}{\emph{Phys. Rev.}
		{\bfseries D101} (2020) 024002}
	[\href{https://arxiv.org/abs/1909.10887}{{\ttfamily 1909.10887}}].
	
	\bibitem{Jenks:2023pmk}
	L.~Jenks, L.~Choi, M.~Lagos and N.~Yunes, \emph{{Parametrized parity violation
			in gravitational wave propagation}},
	\href{https://doi.org/10.1103/PhysRevD.108.044023}{\emph{Phys. Rev. D}
		{\bfseries 108} (2023) 044023}
	[\href{https://arxiv.org/abs/2305.10478}{{\ttfamily 2305.10478}}].
	
	\bibitem{Nishizawa:2018srh}
	A.~Nishizawa and T.~Kobayashi, \emph{{Parity-violating gravity and GW170817}},
	\href{https://doi.org/10.1103/PhysRevD.98.124018}{\emph{Phys. Rev.}
		{\bfseries D98} (2018) 124018}
	[\href{https://arxiv.org/abs/1809.00815}{{\ttfamily 1809.00815}}].
	
	\bibitem{Wang:2020cub}
	Y.-F.~Wang, R.~Niu, T.~Zhu and W.~Zhao, \emph{{Gravitational Wave Implications
			for the Parity Symmetry of Gravity in the High Energy Region}},
	\href{https://doi.org/10.3847/1538-4357/abd7a6}{\emph{Astrophys. J.}
		{\bfseries 908} (2021) 58}
	[\href{https://arxiv.org/abs/2002.05668}{{\ttfamily 2002.05668}}].
	
	\bibitem{Wang:2021gqm}
	Y.-F.~Wang, S.M.~Brown, L.~Shao and W.~Zhao, \emph{{Tests of Gravitational-Wave
			Birefringence with the Open Gravitational-Wave Catalog}},
	\href{https://arxiv.org/abs/2109.09718}{{\ttfamily 2109.09718}}.
	
	\bibitem{Okounkova:2021xjv}
	M.~Okounkova, W.M.~Farr, M.~Isi and L.C.~Stein, \emph{{Constraining
			gravitational wave amplitude birefringence and Chern-Simons gravity with
			GWTC-2}}, \href{https://doi.org/10.1103/PhysRevD.106.044067}{\emph{Phys. Rev.
			D} {\bfseries 106} (2022) 044067}
	[\href{https://arxiv.org/abs/2101.11153}{{\ttfamily 2101.11153}}].
	
	\bibitem{Zhu:2023rrx}
	T.~Zhu, W.~Zhao, J.-M.~Yan, Y.-Z.~Wang, C.~Gong and A.~Wang, \emph{{Constraints
			on parity and Lorentz violations in gravity from GWTC-3 through a
			parametrization of modified gravitational wave propagations}},
	\href{https://doi.org/10.1103/PhysRevD.110.064044}{\emph{Phys. Rev. D}
		{\bfseries 110} (2024) 064044}
	[\href{https://arxiv.org/abs/2304.09025}{{\ttfamily 2304.09025}}].
	
	\bibitem{Callister:2023tws}
	T.~Callister, L.~Jenks, D.~Holz and N.~Yunes, \emph{{A New Probe of
			Gravitational Parity Violation Through (Non-)Observation of the Stochastic
			Gravitational-Wave Background}},
	\href{https://arxiv.org/abs/2312.12532}{{\ttfamily 2312.12532}}.
	
	\bibitem{Zhang:2025kcw}
	B.-Y.~Zhang, T.~Zhu, J.-M.~Yan, J.-F.~Zhang and X.~Zhang, \emph{{Constraining
			parity and Lorentz violations in gravity with future ground- and space-based
			gravitational wave detectors}},
	\href{https://doi.org/10.1103/PhysRevD.111.104012}{\emph{Phys. Rev. D}
		{\bfseries 111} (2025) 104012}
	[\href{https://arxiv.org/abs/2502.04776}{{\ttfamily 2502.04776}}].
	
	\bibitem{Zhang:2022xmm}
	F.~Zhang, J.-X.~Feng and X.~Gao, \emph{{Circularly polarized scalar induced
			gravitational waves from the Chern-Simons modified gravity}},
	\href{https://doi.org/10.1088/1475-7516/2022/10/054}{\emph{JCAP} {\bfseries
			10} (2022) 054} [\href{https://arxiv.org/abs/2205.12045}{{\ttfamily
			2205.12045}}].
	
	\bibitem{Feng:2023veu}
	J.-X.~Feng, F.~Zhang and X.~Gao, \emph{{Scalar induced gravitational waves from
			Chern-Simons gravity during inflation era}},
	\href{https://doi.org/10.1088/1475-7516/2023/07/047}{\emph{JCAP} {\bfseries
			07} (2023) 047} [\href{https://arxiv.org/abs/2302.00950}{{\ttfamily
			2302.00950}}].
	
	\bibitem{Zhang:2023scq}
	F.~Zhang, J.-X.~Feng and X.~Gao, \emph{{Scalar induced gravitational waves in
			symmetric teleparallel gravity with a parity-violating term}},
	\href{https://doi.org/10.1103/PhysRevD.108.063513}{\emph{Phys. Rev. D}
		{\bfseries 108} (2023) 063513}
	[\href{https://arxiv.org/abs/2307.00330}{{\ttfamily 2307.00330}}].
	
	\bibitem{Li:2024fxy}
	T.-C.~Li, T.~Zhu, W.~Zhao and A.~Wang, \emph{{Power spectra and circular
			polarization of primordial gravitational waves with parity and Lorentz
			violations}},
	\href{https://doi.org/10.1088/1475-7516/2024/07/005}{\emph{JCAP} {\bfseries
			07} (2024) 005} [\href{https://arxiv.org/abs/2403.05841}{{\ttfamily
			2403.05841}}].
	
	\bibitem{Moretti:2024fzb}
	T.~Moretti, N.~Bartolo and A.~Greco, \emph{{Breaking parity: the case of the
			trispectrum from chiral scalar-tensor theories of gravity}},
	\href{https://doi.org/10.1088/1475-7516/2025/07/051}{\emph{JCAP} {\bfseries
			07} (2025) 051} [\href{https://arxiv.org/abs/2410.11801}{{\ttfamily
			2410.11801}}].
	
	\bibitem{Feng:2026pcs}
	J.-X.~Feng, J.-Y.~Fang and X.~Gao, \emph{{Circularly polarized gravitational
			waves from parity-violating scalar-tensor theory}},
	\href{https://arxiv.org/abs/2602.07430}{{\ttfamily 2602.07430}}.
	
	\bibitem{Yu:2025fcv}
	Z.-K.~Yu, L.~Liu and T.~Zhu, \emph{{Slowly rotating black hole in chiral
			scalar{\textendash}tensor theory}},
	\href{https://doi.org/10.1140/epjc/s10052-025-14096-y}{\emph{Eur. Phys. J. C}
		{\bfseries 85} (2025) 387}
	[\href{https://arxiv.org/abs/2502.19836}{{\ttfamily 2502.19836}}].
	
	\bibitem{Li:2020xjt}
	M.~Li, H.~Rao and D.~Zhao, \emph{{A simple parity violating gravity model
			without ghost instability}},
	\href{https://doi.org/10.1088/1475-7516/2020/11/023}{\emph{JCAP} {\bfseries
			11} (2020) 023} [\href{https://arxiv.org/abs/2007.08038}{{\ttfamily
			2007.08038}}].
	
	\bibitem{Li:2021wij}
	M.~Li, H.~Rao and Y.~Tong, \emph{{Revisiting a parity violating gravity model
			without ghost instability: Local Lorentz covariance}},
	\href{https://doi.org/10.1103/PhysRevD.104.084077}{\emph{Phys. Rev. D}
		{\bfseries 104} (2021) 084077}
	[\href{https://arxiv.org/abs/2104.05917}{{\ttfamily 2104.05917}}].
	
	\bibitem{Conroy:2019ibo}
	A.~Conroy and T.~Koivisto, \emph{{Parity-Violating Gravity and GW170817 in
			Non-Riemannian Cosmology}},
	\href{https://doi.org/10.1088/1475-7516/2019/12/016}{\emph{JCAP} {\bfseries
			12} (2019) 016} [\href{https://arxiv.org/abs/1908.04313}{{\ttfamily
			1908.04313}}].
	
	\bibitem{Sulantay:2022sag}
	F.~Sulantay, M.~Lagos and M.~Ba{\~n}ados, \emph{{Chiral gravitational waves in
			Palatini-Chern-Simons gravity}},
	\href{https://doi.org/10.1103/PhysRevD.107.104025}{\emph{Phys. Rev. D}
		{\bfseries 107} (2023) 104025}
	[\href{https://arxiv.org/abs/2211.08925}{{\ttfamily 2211.08925}}].
	
	\bibitem{Crisostomi:2017ugk}
	M.~Crisostomi, K.~Noui, C.~Charmousis and D.~Langlois, \emph{{Beyond Lovelock
			gravity: Higher derivative metric theories}},
	\href{https://doi.org/10.1103/PhysRevD.97.044034}{\emph{Phys. Rev.}
		{\bfseries D97} (2018) 044034}
	[\href{https://arxiv.org/abs/1710.04531}{{\ttfamily 1710.04531}}].
	
	\bibitem{Hu:2024hzo}
	Y.-M.~Hu and X.~Gao, \emph{{Parity-violating scalar-tensor theory}},
	\href{https://doi.org/10.1103/PhysRevD.110.064038}{\emph{Phys. Rev. D}
		{\bfseries 110} (2024) 064038}
	[\href{https://arxiv.org/abs/2405.20158}{{\ttfamily 2405.20158}}].
	
	\bibitem{Song:2026ajt}
	J.-J.~Song and X.~Gao, \emph{{Parity-violating spatially covariant gravity at
			total derivative order $d=5$}},
	\href{https://arxiv.org/abs/2607.20317}{{\ttfamily 2607.20317}}.
	
\end{thebibliography}
\end{document}